\documentstyle[12pt]{article}

\newcommand{\mysection}{\setcounter{equation}{0}\section}

\begin{document}
\vskip 0.2cm
\hfill{NIKHEF/95-070} 
\vskip 0.2cm
\hfill{ITP-SB-95-59} 
\vskip 0.2cm
\hfill{INLO-PUB-22/95} 
\vskip 0.2cm
\centerline{\large\bf {Heavy quark coefficient functions  }}
\centerline{\large\bf {at asymptotic values $Q^2 \gg m^2$ }}
\vskip 0.2cm
\centerline {\sc M. Buza \footnote{supported by the Foundation 
for Fundamental Research on Matter (FOM)}}
\centerline{\it NIKHEF/UVA,}
\centerline{\it POB 41882, NL-1009 DB Amsterdam,}
\centerline{\it The Netherlands.}
\vskip 0.2cm 
\centerline {\sc Y. Matiounine and J. Smith}
\centerline{\it Institute for Theoretical Physics,}
\centerline{\it State University of New York at Stony Brook,}
\centerline{\it New York 11794-3840, USA.}
\vskip 0.2cm
\centerline {\sc R. Migneron \footnote{partially supported by the 
Netherlands Organization for Scientific Research (NWO)}}
\centerline{\it Department of Applied Mathematics,}
\centerline{\it University of Western Ontario,}
\centerline{\it London, Ontario, N6A 5B9, Canada.}
\vskip 0.2cm
\centerline {\sc W.L. van Neerven}
\centerline{\it Instituut-Lorentz,}
\centerline{\it University of Leiden,}
\centerline{\it PO Box 9506, 2300 RA Leiden,}
\centerline{\it The Netherlands.}
\vskip 0.2cm
\centerline{December 1995}
\vskip 0.2cm
\centerline{\bf Abstract}
\vskip 0.3cm
In this paper we present the analytic form of the heavy-quark coefficient 
functions for deep-inelastic lepton-hadron scattering in the
kinematical regime $Q^2 \gg m^2$ . Here $Q^2$ and $m^2$ stand for the masses
squared of the virtual photon and heavy quark respectively .
The calculations have been performed up to next-to-leading 
order in the strong coupling constant $\alpha_s$
using operator product expansion techniques. Apart from a check on earlier
calculations, which however are only accessible via large
computer programs, the asymptotic forms of the coefficient functions
are useful for charm production at HERA when the condition 
 $Q^2 \gg m_c^2$ is satisfied. Furthermore the 
analytical expressions can also be used when one applies the variable
heavy flavour scheme up to next-to-leading order in $\alpha_s$.
\vfill
\newpage
\mysection{Introduction}
The study of deep inelastic electroproduction has led to 
important information on the structure of the proton. This information
is extracted from the structure functions $F_2(x,Q^2)$ and $F_L(x,Q^2)$
which appear in the cross section for deep inelastic lepton-hadron scattering.
Here $x$ denotes the Bjorken scaling variable and $Q^2$ is the mass squared of
the virtual photon exchanged between the lepton and the hadron. In the 
framework of perturbative Quantum Chromodynamics (QCD) these 
structure functions can be described by a convolution of parton densities 
with coefficient functions.
The latter are calculable order by order in perturbation theory. However the
parton densities cannot be computed yet since they are of a nonperturbative
origin and have to be extracted from the data. Starting from the late
sixties these densities have been obtained from many experiments. 
Until recently the analysis was 
carried out in the kinematical range $0.01<x<0.95$ and 
$Q^2<300$ (GeV/c)$^2$. However since the advent of 
the HERA accelerator the kinematical region has been
extended to much smaller values of $x$ ($x > 10^{-4}$) and much larger 
values of $Q^2$ ($Q^2 < 2\times10^4$ (GeV/c)$^2$). 
The most recent results come from the H1  and 
ZEUS experiments at HERA, see \cite {H1} and \cite{ZEUS} respectively.

The low-$x$ region is of great experimental as well as theoretical interest.
The structure function $F_2(x,Q^2)$ rises very steeply when 
$x \rightarrow 0$ 
which can be mainly attributed to a corresponding increase in the
gluon density. Therefore this density is a very important 
issue in the investigation of the small-$x$ structure
of the proton. Since the gluon density appears together with the other parton 
densities in most cross sections, one has to look for those specific 
reactions in which it plays a dominant role so that it can be isolated 
from its partners. One of these processes is extrinsic charm 
production in deep inelastic electron-proton scattering. Here the dominant 
production mechanism is represented by the photon-gluon fusion process which
is indeed the only one in the Born approximation. Next-to-leading order (NLO) 
calculations \cite{lrsn1}, to which also other processes contribute, reveal 
that this picture remains unaltered. 

Apart from the interest in the gluon density, charm production 
also revived the important issue of how to treat the charm
quark in deep inelastic scattering processes. Here one can distinguish between
intrinsic \cite{bhmt} and extrinsic charm production. In the former 
case the charm is 
considered to be a part of the hadronic wave function and it is described by a 
parton density in the hadron like the other light flavours ($u$,$d$,$s$) and 
the gluon ($g$). However this prescription is only correct 
if the charm quark can
be treated as a massless particle which is certainly not the case at 
small-$Q^2$ values where threshold effects become important. In this region 
the charm quark has to be treated as a massive particle. On the other hand when
$Q^2 \gg m_c^2$ ,where $m_c$ is the charm mass, extrinsic charm production via 
the photon-gluon fusion process and its higher order QCD corrections reveal
large logarithms of the type $\alpha_s^k \ln^k(Q^2/m_c^2)$. These large
corrections bedevil the perturbation series and have therefore to be resummed 
via the renormalization group equations. This resummation entails the
definition of a charm parton density in the hadron. Although a recent
investigation \cite{grs} shows that the above logarithms lead to a rather 
stable cross section for charm production with respect to variations in the
factorization and renormalization scale, the size of these large corrections
warrants a special treatment. This is provided by the so-called variable
flavour scheme (VFS) \cite{cwz}, \cite{ctt}. 
In this scheme the treatment of the charm
depends on the values chosen for $Q^2$. At low $Q^2$-values the deep inelastic
structure functions are described by the light parton densities 
($u$,$d$,$s$,$g$). The charm contribution is given 
by the photon-gluon fusion process and its
higher order QCD corrections. At large $Q^2$ the charm is treated in the same 
way as the other light quarks and it is represented by a charm parton density
in the hadron, which evolves in $Q^2$. 
In the intermediate $Q^2$ region one has to make a smooth 
connection between the two different prescriptions. In \cite{aot},
\cite{or} and \cite{kls} this was done by adding and subtracting 
certain mass factorization terms. Notice that the above
considerations also apply to bottom production when $Q^2$ gets extremely large.

Up to now the VFS-scheme has only been applied to 
heavy flavour electroproduction using the Born approximation to the
photon-gluon fusion process \cite{aot}, \cite{or} and \cite{kls}. 
It is our aim to extend this scheme to 
next-to-leading order in $\alpha_s$. For that purpose we will 
calculate the full two-loop
operator matrix elements containing one heavy quark loop. This calculation
provides us with the terms containing the  
logarithms $\ln^k(\mu^2/m_c^2)$ which have to be
subtracted from the charm cross section so that the final result becomes
independent of the charm mass $m_c$. Here $\mu$ denotes the operator 
renormalization scale which can be identified with the factorization scale.
Using the two-loop operator matrix elements up to non-logarithmic terms and the 
NLO light parton coefficient functions in \cite{zn}, we can 
construct an analytic
form of the NLO heavy quark coefficient functions in the 
limit $Q^2 \gg m_c^2$.
We will refer to these results as the asymptotic heavy quark
coefficient functions.
These expressions serve as a check on the exact calculation in \cite{lrsn1} 
which is only available in a large computer program
involving numerical integrations over several variables. 
Furthermore it enables us 
to see at which $Q^2$-values the asymptotic heavy quark coefficient functions
coincide numerically
with the exact ones, which gives an indication when the charm quark can
be treated as a massless quark.

The content of this paper can be summarized as follows. In section 2 we 
introduce our notations and give an outline how the heavy quark coefficient 
functions can be determined in the asymptotic limit 
$Q^2 \gg m_c^2$. In section 3
we present the calculation of the full two-loop operator matrix element needed
for the computation of the asymptotic form of the heavy quark coefficient
function. The latter will be calculated in section 4. Finally in section 5
we will show at which $Q^2$-values the asymptotic form coincides with the 
exact one given in \cite{lrsn1}. In Appendix A an exact analytic expression 
for the heavy quark coefficient function, 
which is valid for any $Q^2$ and $m_c^2$, 
is presented in the case of the Compton subprocess. 
An important trick how
to compute an operator matrix element with five different propagators is given
in Appendix B. The long formulae obtained for the full 
operator matrix elements and the asymptotic heavy quark coefficient 
functions are presented in Appendices C and D respectively.
\vfill
\newpage
\mysection{Heavy flavour coefficient functions}
In the section we will show the connection between the
heavy flavour coefficient functions  computed
in the asymptotic limit $Q^2 \gg m^2$ and the operator matrix elements (OME's). 
Here $Q^2$ denotes the mass squared of the virtual photon with momentum
$q$ ($q^2 = -Q^2 < 0$) and $m$ stands for the heavy flavour mass.
The variable $x$ is defined by $x = Q^2/2p\cdot q$ (Bjorken scaling variable)
where $p$ stands for the momentum of the proton. The OME's 
arise when the local operators, which show up in the operator 
product expansion (OPE) of two electromagnetic currents, are sandwiched
between the proton states indicated by $|P>$. 
In the limit that $x$ is fixed and 
$Q^2\gg M^2$, where $M$ denotes the mass
of the proton,
this OPE dominates the integrand of the 
hadronic structure tensor $W_{\mu \nu}$, which is defined by
\begin{eqnarray}
W_{\mu \nu} (p,q) = {{1}\over{4 \pi}} \int \, d^4z e^{iq\cdot z}
< P |[ J_\mu(z), J_\nu(0)] |P> \,.
\end{eqnarray}
This structure tensor arises when one computes the cross section
for deep inelastic electroproduction of heavy flavours
\begin{eqnarray}
e^-(\ell_1)+P(p)\rightarrow e^-(\ell_2)+Q(p_1)(\ \overline{Q}(p_2)\ )
+'X' \,,
\end{eqnarray}
provided the above reaction is completely inclusive with respect to
the hadronic state $'X'$ as well as the heavy flavours $Q(\ \overline{Q}\ )$.
When the virtuality $Q^2\ (\ q=\ell_1-\ell_2\ )$ of the exchanged photon
is not too large $(\ Q^2\ll M_Z^2\ )$ the reaction in $(2.2)$ is dominated
by the one-photon exchange mechanism and we can neglect any weak effects
caused by the exchange of the Z-boson. In this case $W_{\mu\nu}(p,q)$ 
in $(2.1)$ can be written as
\begin{eqnarray}
 && W_{\mu\nu}(p,q)=\frac{1}{2x}\Big(g_{\mu\nu}
-\frac{q_\mu q_\nu}{q^2}\Big)F_L(x,Q^2)
+\Big(p_\mu p_\nu-\frac{p.q}{q^2}(p_\mu q_\nu+p_\nu q_\mu)
\nonumber \\ && \qquad \qquad\qquad
+g_{\mu\nu}\frac{(p.q)^2}{q^2}\Big)\frac{F_2(x,Q^2)}{p.q} ~ .
\end{eqnarray}
The above formula follows from Lorentz covariance of $W_{\mu\nu}(p,q)$,
parity invariance and the conservation of the electromagnetic current
$J_\mu$. Besides the longitudinal structure function $F_L(x,Q^2)$ and the
structure function $F_2(x,Q^2)$ one can also define the 
transverse structure function $F_1(x,Q^2)$ which, however, 
depends on the two previous ones and is given by
\begin{eqnarray}
F_1(x,Q^2)=\frac{1}{2x}\Big[ F_2(x,Q^2)-F_L(x,Q^2) \Big] ~  .
\end{eqnarray}
The above structure functions  show up in the deep inelastic
cross section of the process in $(2.2)$ provided we integrate over
the whole final state
\begin{eqnarray}
\frac{d^2\sigma}{dx dy}=\frac{2\pi\alpha^2}{(Q^2)^2}S\Big[\{1+(1-y)^2\}
F_2(x,Q^2)-y^2 F_L(x,Q^2)\Big] \,,
\end{eqnarray}
where $S$ denotes the square of the c.m. energy of the electron proton
system and the variables $x$ and $y$ are defined ( see above $(2.1)$\ ) as
\begin{eqnarray}
x=\frac{Q^2}{2p\cdot q}\quad (0<x\le 1)\quad ,\quad  y
=\frac{p\cdot q}{p\cdot \ell_1}
\quad (0<y<1) \,,
\end{eqnarray}
with 
\begin{eqnarray}
-q^2=Q^2=xyS  ~.
\end{eqnarray}
In the QCD improved parton model the heavy flavour contribution
to the hadronic structure functions
can be expressed as integrals over the partonic scaling 
variable $z= Q^2/(s+Q^2)$, 
where $s$ is the square of the photon-parton centre-of-mass 
energy $(s \ge 4m^2)$. 
This yields the following result 
\begin{eqnarray}
 &&F_i(x,Q^2,m^2)=x\int^{z_{max}}_x \frac{dz}{z}\Big[\frac{1}{n_f}
\sum^{n_f}_{k=1}e_k^2 \Big\{\Sigma\Big(\frac{x}{z},\mu^2\Big)
L^{\rm S}_{i,q}(z,Q^2,m^2,\mu^2)
\nonumber \\ && \qquad\qquad
+G\Big(\frac{x}{z},\mu^2\Big) L_{i,g}(z,Q^2,m^2,\mu^2)\Big\}
+\Delta\Big(\frac{x}{z},\mu^2\Big)
L^{\rm NS}_{i,q}(z,Q^2,m^2,\mu^2)\Big]
\nonumber \\ && \qquad\qquad
+x~e_H^2\int^{z_{max}}_{x}\frac{dz}{z}\Big\{\Sigma\Big(\frac{x}{z},\mu^2\Big)
H_{i,q}(z,Q^2,m^2,\mu^2)
\nonumber \\ && \qquad\qquad
+G\Big(\frac{x}{z},\mu^2\Big)H_{i,g}(z,Q^2,m^2,\mu^2)\Big\} \,,
\end{eqnarray}
where $i=2,L$ and 
the upper boundary of the integration is given by
$z_{max}={Q^2}/(4 m^2 + Q^2)$. The function $G(z,\mu^2)$ stands for the gluon 
density. The singlet combination of the quark densities is defined by 
\begin{eqnarray}
\Sigma(z,\mu^2)=\sum^{n_f}_{i=1}\Big(f_i(z,\mu^2)+\overline{f}_i(z,\mu^2)
\Big) \,,
\end{eqnarray}
where $f_i$ and $\overline{f}_i$ stand for the light quark and anti-quark
densities of species $i$ respectively. The non-singlet combination of
the quark densities is given by
\begin{eqnarray}
\Delta(z,\mu^2)=\sum^{n_f}_{i=1}\Big(e_i^2
-\frac{1}{n_f}\sum^{n_f}_{k=1}e_k^2\Big)
\Big(f_i(z,\mu^2)+\overline{f}_i(z,\mu^2)\Big) ~ .
\end{eqnarray}
In the above expressions the charges of the light quark and the heavy quark
are denoted by $e_i$ and $e_H$ respectively. Furthermore, $n_f$ stands
for the number of light quarks and $\mu$ denotes the mass factorization
scale, which we choose to be equal to the  renormalization scale.
The latter shows up in the running coupling constant defined by
$\alpha_s(\mu^2)$. Like the parton densities the heavy quark coefficient
functions $L_{i,j}$ $(i=2,L$; $j=q,g)$ can also be divided 
into singlet and non-singlet parts which are indicated by the
superscripts S and NS in eq. $(2.8)$.

The distinction between the heavy quark coefficient functions $L_{i,j}$ and 
$H_{i,j}$ can be traced back to the different photon-parton 
production procesess from which they originate. The functions
$L_{i,j}$ are attributed to the reactions where the virtual photon
couples to the light quark, whereas $H_{i,j}$ originates from the reactions
where the virtual photon couples to the heavy quark. 
This explains why there is
only a singlet part for $H_{i,j}$ and that $L_{i,j}$, $H_{i,j}$ in eq. $(2.8)$
are multiplied by $e_i^2$, $e_H^2$ respectively. 
In \cite{lrsn1} the heavy quark coefficient
functions have been calculated up to next-to-leading order(NLO). In the Born
approximation (first order of $\alpha_s$ or LO) one has the photon-gluon fusion
process
\begin{eqnarray}
\gamma^*(q) + g(k_1) \rightarrow Q(p_1) + \overline{Q}(p_2) \,,
\end{eqnarray}
which leads to the lowest order contribution to $H_{i,g}$ denoted
by $H_{i,g}^{(1)}$. The next order is obtained by including the virtual gluon
corrections to process $(2.11)$ and the gluon bremsstrahlung process
\begin{eqnarray}
\gamma^*(q) + g(k_1) \rightarrow g(k_2) + Q(p_1) + \overline{Q}(p_2) \,,
\end{eqnarray}
both of which contribute to the second order term in $H_{i,g}$ denoted
by $H_{i,g}^{(2)}$. In addition to the above reaction we also have the
subprocess where the gluon in $(2.12)$ is replaced by a light (anti-)
quark, i. e.
\begin{eqnarray}
\gamma^*(q) +q(\overline{q})(k_1)\rightarrow q(\overline{q})(k_2)
+ Q(p_1) + \overline{Q}(p_2) ~ .
\end{eqnarray}
This process has however two different production mechanisms. The first one
is given by the Bethe-Heitler process (see figs. 5a,b in \cite{lrsn1}) 
and the second one can be attributed to 
the Compton reaction (see figs. 5c,d in \cite{lrsn1}).
In the case of the Bethe-Heitler process the virtual photon couples to
the heavy quark and therefore this reaction contributes to $H_{i,q}$.
This second order contribution will be denoted by $H_{i,q}^{(2)}$.
In the Compton reaction the virtual photon couples to the light (anti-)
quark and its contribution to  $L_{i,q}^{\rm NS}$ will be 
denoted by  $L_{i,q}^{{\rm NS},(2)}$.
Since $L_{i,q}^{\rm S}$ can be written as $L_{i,q}^{\rm S}=L_{i,q}^{\rm NS} + 
L_{i,q}^{\rm PS}$, $L_{i,q}^{{\rm NS},(2)}$ also contributes to the singlet part
of the coefficient function.
In general the heavy quark coefficient functions 
are expanded in $\alpha_s$ as follows 
\begin{eqnarray}
H_{i,g}(z,Q^2,m^2,\mu^2)=\sum^{\infty}_{k=1}\Big(\frac{\alpha_s}{4\pi}
\Big)^k H_{i,g}^{(k)}(z,Q^2,m^2,\mu^2) ~ ,
\end{eqnarray}
\begin{eqnarray}
H_{i,q}(z,Q^2,m^2,\mu^2)=\sum^{\infty}_{k=2}\Big(\frac{\alpha_s}{4\pi}
\Big)^k H_{i,q}^{(k)}(z,Q^2,m^2,\mu^2) ~,
\end{eqnarray}
\begin{eqnarray}
L^{r}_{i,j}(z,Q^2,m^2,\mu^2)=\sum^{\infty}_{k=2}\Big(\frac{\alpha_s}{4\pi}
\Big)^k L_{i,j}^{r,(k)}(z,Q^2,m^2,\mu^2) \,,
\end{eqnarray}
where $r = $ NS,S.
Finally we want to make the remark
that there are no interference terms between the Bethe-Heitler and
Compton reactions in $(2.13)$ if one integrates over all final state momenta.

The complexity of the second order heavy quark coefficient functions prohibits
publishing them in an analytic form, except for $L_{i,q}^{{\rm NS},(2)}$,
which is given in Appendix A, so that they are only
available in large computer programs \cite{lrsn1}, involving two-dimensional
integrations. 
To shorten the long running time for these programs
we have previously tabulated the coefficient functions in the form of a two 
dimensional array in the variables $\eta$ and $\xi$ in 
a different computer program \cite{rsn1}. These variables are
defined by ( using $s = (q + k_1)^2$, $z = Q^2/(s+Q^2)$,
and $\eta = (s - 4m^2)/(4m^2)$)
\begin{eqnarray}
\eta=\frac{(1-z)}{4z}\xi -1\qquad,\qquad \xi=\frac{Q^2}{m^2}  ~.
\end{eqnarray}
This new program has shortened the computation of the charm structure functions 
$F_i(x,Q^2,m_c^2)$ considerably as one only requires one integral over
the variable $z$ in (2.8), therefore making our results for the NLO
corrections more amenable for phenomenological applications. 
However, when $Q^2\gg m^2$ it is possible
to get complete analytic forms for the heavy quark coefficient functions
which are similar to the ones presented for the light quark and gluon
coefficient functions given in \cite{zn}. To get the analytic form in the above
asymptotic regime one can follow two approaches. The first one is to go back
to the original calculation of the exact coefficient functions in 
\cite{lrsn1} and repeat the computation of the Feynman 
graphs and the phase space integrals
in the limit $Q^2\gg m^2$. An example of such a calculation can be found 
in \cite{bnb}, where all photonic corrections to the initial state of the
process $e^-+e^+\rightarrow \mu^-+\mu^+$ in the limit $S\gg m^2_e$
were computed. 
However, this procedure is still quite complicated because one cannot neglect
the fermion masses at too premature a stage which 
results in rather messy calculations. 
Fortunately, as one can find in \cite{bnb}, there exists an
alternative method which we will use for the heavy quark coefficient
functions.

In the limit $Q^2\gg m^2$ the heavy quark coefficient functions behave
logarithmically as
\begin{eqnarray}
H_{i,j}^{(k)}(z,Q^2,m^2,\mu^2)=\sum^k_{l=1}a^{(k,l)}_{i,j}
\Big(z,\frac{\mu^2}{m^2}\Big)\ln^l\frac{Q^2}{m^2}  ~,
\end{eqnarray}
with a similar expression for L$_{i,j}^{(k)} $.
As has been already mentioned in the introduction these large logarithms
$\ln^l(Q^2/m^2)$ dominate the radiative corrections. This is in particular
the case for charm production in the large $Q^2$ region which is accesible
to HERA experiments. The above large logarithms also arise when $Q^2$ is kept
fixed and $m^2\rightarrow 0$ so that they originate from collinear 
singularities. These collinear divergences can be removed via mass 
factorization. The latter proceeds in the following way. In the non-singlet
case we have
\begin{eqnarray}
C^{\rm NS}_{i,q}\Big(\frac{Q^2}{\mu^2},n_f\Big)+L^{\rm NS}_{i,q}
\Big(\frac{Q^2}{m^2},
\frac{\mu^2}{m^2}\Big)=\Gamma^{\rm NS}_{qq}\Big(\frac{\mu^2}{m^2}\Big)
\otimes C^{\rm NS}_{i,q}
\Big(\frac{Q^2}{\mu^2},n_f+1\Big)  ~.
\end{eqnarray}
For the singlet case the mass factorization becomes
\begin{eqnarray}
 &&C^{\rm S}_{i,l}\Big(\frac{Q^2}{\mu^2},n_f\Big)+L^{\rm S}_{i,l}
\Big(\frac{Q^2}{m^2},
\frac{\mu^2}{m^2}\Big)+H_{i,l}\Big(\frac{Q^2}{m^2},\frac{\mu^2}{m^2}\Big)=
\nonumber \\ && \qquad\qquad
\sum_k\Gamma^{\rm S}_{kl}\Big(\frac{\mu^2}{m^2}\Big)\otimes C^{\rm S}_{i,k}
\Big(\frac{Q^2}{\mu^2},n_f+1\Big) \,,
\end{eqnarray}
with $k,l=q,g$. Notice that in the above expression we have suppressed the
$z$-dependence for simplicity. The convolution symbol is defined by
\begin{eqnarray}
\Big(f\otimes g\Big)(z)=\int_0^1 dz_1\int_0^1 dz_2  ~
\delta(z-z_1z_2)f(z_1)g(z_2)  ~.
\end{eqnarray}
The quantities $\Gamma_{kl}$ and $C_{i,k}$ ($k,l=q,g$) which appear
in the above equation stand for the transition functions and the light
parton coefficient functions respectively. Notice that all mass dependence is 
transfered to the transition function $\Gamma_{kl}$. 
The removal of the logarithmic
terms $\ln^l(\mu^2/m^2)$  from the heavy flavour coefficient functions 
leads to an enhancement by one of the number of light flavours $n_f$ 
in the light parton coefficient functions $C_{i,k}$ - see eqs.$(2.19)$,
$(2.20)$. The latter have been calculated up to second order in
$\alpha_s$ in \cite{zn}. 
Since these coefficient functions do not depend on the way
one has regularized the collinear divergences in the parton cross sections
(for a discussion see \cite{hn}) one can also use them 
in equations $(2.19)$,
$(2.20)$. Now if one also knows the transition functions $\Gamma_{kl}$
one can reconstruct the asymptotic behavior of the heavy flavour
coefficient functions $L_{i,k}$ and $H_{i,k}$. This is possible
because the transition functions $\Gamma_{kl}$ are identical to 
the operator matrix elements (OME) denoted by $A_{kl}$. The latter
appear in the operator product expansion (OPE) 
of the commutator of the two electromagnetic currents in (2.1)
near the lightcone. Suppressing the Lorentz index of the electromagnetic
current the expansion can be written as
\begin{eqnarray}
\lim_{z^2\rightarrow 0}[J(z),J(0)]
=\sum_k\sum_m C^{(m)}_k(z^2)z_{\mu_1}
\cdots z_{\mu_m}O^{\mu_1\cdots\mu_m}_k(0) \,,
\end{eqnarray}
where the distributions $C^{(m)}_k(z^2)$ $(k=q,g)$ are 
the Fourier transforms of the light
parton coefficient functions $C^{(m)}_k(Q^2/\mu^2)$ defined in $(2.19)$,
$(2.20)$. Notice that the latter has been Mellin transformed according to
\begin{eqnarray} 
C^{(m)}_k\Big(\frac{Q^2}{\mu^2}\Big)=\int_0^1 dy \, y^{m-1}
 C_k\Big(y,\frac{Q^2}{\mu^2}\Big) ~.
\end{eqnarray}
In the above expressions the index $i$ ($i=2,L$) 
in $C^{(m)}_{i,k}$ has been omitted 
since we suppressed the Lorentz index of the current $J_\mu(z)$.
Inclusion of the latter index implies that one has two independent
structure functions $F_2$ and $F_L$. The quantity $m$ refers to the spin 
of the local operators $O_k^{\mu_1\cdots\mu_m}$ which appear 
in the OPE $(2.22)$.
The latter are given by the non-singlet quark operator
\begin{eqnarray} 
O_{q,r}^{\mu_1\cdots\mu_m}(x)=\frac{1}{2}i^{m-1}S\Big[\overline{\psi}(x)
\gamma^{\mu_1}D^{\mu_2}\cdots D^{\mu_m}\frac{\lambda_r}{2}\psi(x)\Big]
+{\rm trace\  terms} \,,
\end{eqnarray}
and the singlet operators
\begin{eqnarray} 
O_q^{\mu_1\cdots\mu_m}(x)=\frac{1}{2}i^{m-1}S\Big[\overline{\psi}(x)
\gamma^{\mu_1}D^{\mu_2}\cdots D^{\mu_m}\psi(x)\Big]
+{\rm trace\  terms}  ~,
\end{eqnarray}
\begin{eqnarray} 
&&O_g^{\mu_1\cdots\mu_m}(x)=\frac{1}{2}i^{m-2}S\Big[ F_{\alpha}^{a,\mu_1}(x)
D^{\mu_2}\cdots D^{\mu_{m-1}} F^{a,\alpha\mu_m}(x)\Big]
+{\rm trace\  terms}  ~.
\nonumber \\ &&
\end{eqnarray}
Here $S$ denotes the symmetrization of the operators in their Lorentz indices
$\mu_i$ and the trace terms are needed to make them traceless.
The $\lambda_r$ in $(2.24)$ represent the generators of the flavour algebra,
whereas the index $a$ in $(2.26)$ is the colour index.
The objects $\psi(x)$, $F^a_{\mu\nu}$ represent the quark field
and the gluon field tensor respectively and $D_{\mu}$ denotes the covariant
derivative.

The OPE expansion in $(2.22)$ can be applied in the limit $Q^2\gg m^2$ 
and fixed $x$ 
so that the integrand in $(2.1)$ gets its dominant contribution from
the lightcone behavior of the current-current commutator. Inserting $(2.22)$
in $(2.1)$ and replacing the hadron state $|P>$ by a light quark or gluon 
one gets the relations
\begin{eqnarray} 
&&\tilde {\cal F}^{\rm NS}_{i,q}\Big({{Q^2}\over{\mu^2}}, \epsilon\Big)
+ \tilde L^{\rm NS}_{i,q}\Big({{Q^2}\over{m^2}},
 {{\mu^2}\over{m^2}},\epsilon\Big)
= \tilde A^{\rm NS}_{qq}\Big({{\mu^2}\over{m^2}}, \epsilon\Big) \otimes 
C^{\rm NS}_{i,q}\Big({{Q^2}\over{\mu^2}}\Big)  ~,
\end{eqnarray}
\begin{eqnarray}
&&\tilde {\cal F}^{\rm S}_{i,l}\Big({{Q^2}\over{\mu^2}}, \epsilon\Big)
+\tilde L^{\rm S}_{i,l}\Big({{Q^2}\over{m^2}},{{\mu^2}\over{m^2}},\epsilon\Big)
+\tilde H_{i,l}\Big({{Q^2}\over{m^2}},{{\mu^2}\over{m^2}},\epsilon\Big)
\nonumber  \\ && \qquad \qquad \qquad
= \tilde A^{\rm S}_{kl}\Big({{\mu^2}\over{m^2}},\epsilon\Big) \otimes 
C^{\rm S}_{i,k}\Big({{Q^2}\over{\mu^2}}\Big)  ~,
\end{eqnarray}
where $i=2,L$; $k,l=q,g$.
Since the operators in (2.22) are already renormalized,
the above expressions can be only collinearly divergent.
The collinear divergences due to the presence of light partons
are regularized by the method of $n$-dimensional regularization.
They are indicated by the pole terms $(1/\epsilon)^i$ ($\epsilon = n-4$).
However the collinear divergences due to the heavy quarks are
regularized by the mass $m$ which shows up in the form of logarithms of the
type $\ln^i(Q^2/m^2)$, $\ln^i(\mu^2/m^2)$. 
The objects $\tilde {\cal F}_{i,k}$ 
denote the light parton structure functions. The 
OME's $\tilde A_{kl}$ are defined by
\begin{eqnarray}
\tilde A_{kl}\Big({{\mu^2}\over{m^2}},\epsilon\Big) = <l|O_k|l>  ~,
\end{eqnarray}
where $|l>$ is a light quark or a gluon state and all quantities
depending on the pole terms are indicated by a tilde.
If the latter are removed via mass factorization one obtains 
expressions (2.19), (2.20) which implies the following identification
\begin{eqnarray}
\Gamma^{\rm NS}_{qq}\Big({{\mu^2}\over{m^2}}\Big) = A^{\rm NS}_{qq}
\Big({{\mu^2}\over{m^2}}\Big) ~,
\end{eqnarray}
\begin{eqnarray}
\Gamma^{\rm S}_{kl}\Big({{\mu^2}\over{m^2}}\Big) = A^{\rm S}_{kl}
\Big({{\mu^2}\over{m^2}}\Big) ~.
\end{eqnarray}
For the computation of the asymptotic behaviour of the heavy quark coefficient
functions corresponding to the processes (2.11)-(2.13) one needs the
following quantities. For processes (2.11) and (2.12) we have to calculate the
one-loop OME $A^{(1)}_{Qg}$
and the two-loop OME $A^{(2)}_{Qg}$ respectively.
They are represented by the Feynman graphs in fig.1 and fig.2.
The Bethe-Heitler process given by reaction (2.13) corresponds
to the two-loop OME $A^{{\rm PS},(2)}_{Qq}$ 
in fig.3 whereas the Compton process
(2.13) is related to $A^{{\rm NS},(2)}_{qq}$
with the two-loop OME's in fig.4.
The calculation of the graphs in figs. 1-4 and the derivation
of the asymptotic form of the heavy
quark coefficient functions will be the aim of the next two sections.
\vfill
\newpage
\mysection{Calculation of the two-loop operator matrix elements}
Before presenting the results of our calculation of the OME's
we will first derive the general structure of the OME's discussed in the last
section. If we insert the OPE (2.22) into
the structure tensor $W_{\mu\nu}$ in (2.1) the OME's which are derived from the
Feynman graphs in figs.1-4 will be computed in the forward direction.
The latter means that the momentum leaving the operator vertex equals zero.
Further if one puts the momentum, indicated by $p$, of the external light quark
and gluon off-shell ($p^2 < 0 $) only ultraviolet
(UV) singularities appear in the OME's. Using this off-mass-shell 
assignment one can express the renormalized 
as well as the unrenormalized OME's into the
renormalization group coefficients as is done in \cite{mn}.
However for our computations we have to put the external momentum on-shell
($p^2=0$) so that the OME's turn into genuine S-matrix elements.
This mass assignment implies that in addition to UV divergences one
also encounters collinear (C) divergences which originate from the coupling
of the external on-shell massless quanta to internal massless quanta. 
In the computation of the Feynman graphs both types of divergences will
be regularized using the technique of $n$-dimensional regularization.
However since both singularities in the OME's will manifest themselves 
in the form of pole terms of the type $\epsilon^{-k}$ ($\epsilon = n-4$)
it is very hard to trace back their origins. Nevertheless one can
express the OME's into the renormalization group coefficients in a 
similar way as has been derived for the off-shell case in \cite{mn}.
Where possible we will make a distinction between UV-pole terms
and C-pole terms, which are indicated by
$\epsilon_{UV}^{-k}$ and $\epsilon_{C}^{-k}$ respectively , and
identify them $\epsilon_{UV}=\epsilon_C$ when it is appropriate.

In the subsequent part of this section we will construct the OME's,
corresponding to the graphs in figs.1-4, in such a way that the
coefficients of the pole terms are given by the renormalization group.
These coefficients are products of the terms appearing in the beta-function
and the AP-splitting functions (anomalous dimensions) \cite{glap}. 
The purpose of this
presentation is threefold. First we need these coefficients for the
construction of the heavy quark coefficient functions in the next section.
Second since the renormalization group coefficients are known in the literature
we can predict the residues of the pole terms so that these expressions serve 
as a check on our calculations. Third it is much easier to show the 
renormalization and mass factorization for the algebraic expressions,
which are short, than for the analytic formulae in our 
calculations because the latter are rather long.

The OME's $A_{ij}$ can be expanded in a perturbation series as follows
\begin{eqnarray}
A_{ij} = \sum_{k=0}^{\infty} \Big( \frac{\alpha_s}{4\pi} \Big)^{k}
A_{ij}^{(k)} \,.
\end{eqnarray}
In the following discussion we distinguish three different types
of OME's. First we have the unrenormalized ones indicated by $\hat A_{ij}$.
They contain UV-as well as C-singularities. Second after renormalization
the UV-divergences are removed and we are left by the OME's defined by
$\tilde A_{ij}$ which still contain C-divergences. The latter have to be
removed via mass factorization so that the $\tilde A_{ij}$ turn into the 
finite OME's indicated by $A_{ij}$. Notice that the expansion in (3.1)
holds for all three different types of OME's.

The renormalization of $\hat A_{ij}$ proceeds in three steps.
First we will perform mass renormalization for which we choose
the on-mass-shell scheme. 
This implies that the bare mass $\hat m$, which occurs in 
$\hat A_{ij}$ has to be replaced by
\begin{eqnarray}
\hat m = m ( 1 + \frac{\hat \alpha_s}{4\pi} \delta m ) \quad, \quad
\delta m = C_F S_\epsilon \Big(\frac{m^2}{\mu^2}\Big)^{\epsilon/2}
\Big\{ \frac{6}{\epsilon_{\rm UV}} - 4 \Big\} \,,
\end{eqnarray}
so that the mass renormalized $\hat A_{ij}$ reads up to order $\alpha_s^2$
\begin{eqnarray}
\hat A_{ij} = \hat A_{ij}^{(0)} + 
\big( \frac{\hat \alpha_s}{4 \pi}\Big) \hat A_{ij}^{(1)}
+ \big( \frac{\hat \alpha_s}{4 \pi}\Big)^2
\Big\{ \hat A_{ij}^{(2)} 
+ \delta m \frac{d}{dm} \hat A_{ij}^{(1)} \Big\}\,.
\end{eqnarray}
Notice that the zeroth order term 
$\hat A_{ij}^{(0)} = \delta_{ij}$ is mass independent.
In the above perturbation series the quantities $\mu^2$ and
$S_\epsilon$ are artefacts of $n$-dimensional regularization. The mass
parameter $\mu$ originates from the dimensionality
of the gauge coupling constant $g$ ($\alpha_s = g^2/4\pi$) in
$n$-dimensions and should not be confused with the renormalization
and mass-factorization scale. However if one only subtracts the pole-terms
like in the $\overline{\rm MS}$-scheme the mass parameter $\mu$
turns into the afore-mentioned scales. The spherical factor 
$S_\epsilon$ is defined as
\begin{eqnarray}
S_\epsilon = \exp \Big\{ \frac{\epsilon}{2} (\gamma_E - \ln 4\pi) \Big\}\,,
\end{eqnarray}
and the colour factor is
\begin{eqnarray}
C_F = \frac{ N^2 - 1}{2N} \,,
\end{eqnarray}
in SU(N). Further $\hat \alpha_s$ denotes the 
bare coupling constant which will be renormalized as follows
\begin{eqnarray}
\hat\alpha_s = \alpha_s(\mu^2) \Big( 1 + \frac{\alpha_s(\mu^2)}{4\pi}
\delta \alpha_s \Big) \,,
\end{eqnarray}
\begin{eqnarray}
\delta \alpha_s = S_{\epsilon} \Big\{ 
\frac{2\beta_0}{\epsilon_{\rm UV}}
+ \frac{2\beta_{0,H}}{\epsilon_{\rm UV}}
\Big( \frac{m^2_H}{\mu^2} \Big) ^{\epsilon/2}
\Big( 1 + \frac{1}{8} \zeta(2) \epsilon_{\rm UV}^2 \Big)\Big\} \,.
\end{eqnarray}
Here a summation over the heavy quarks $H ( H=c,b,t)$ is understood
and $\zeta(2) = \pi^2/6 $ is the Riemann zeta-function.
Further $\beta_0$ is the lowest order term in the series expansion
of the beta-function, which up to two-loop order, is given by
\begin{eqnarray}
\beta(g) = - \beta_0 \frac{g^3}{16\pi^2} - \beta_1 \frac{g^5}{(16\pi^2)^2}\,,
\end{eqnarray}
where
\begin{eqnarray}
\beta_0 = \frac{11}{3} C_A - \frac{4}{3} T_f n_f \,,
\nonumber
\end{eqnarray}
and
\begin{eqnarray}
\beta_1 = \frac{34}{3} C_A^2 - 4C_F T_f n_f -  \frac{20}{3} C_A T_f n_f \,.
\nonumber
\end{eqnarray}
Here $C_A$ and $T_f$ denote colour factors of SU(N) 
\begin{eqnarray}
C_A = N \quad , \quad T_f = \frac{1}{2} \,,
\end{eqnarray}
and $n_f$ denotes the number of light flavours which enter via
the fermion-loop contributions to the gluon self-energy.
However besides the light quarks also the heavy quarks with mass
$m_H$ contribute to the renormalized coupling constant $\alpha_s$.
This contribution is indicated in (3.7) by $\beta_{0,H} = - 4T_f/3$ and
$m_H \ge  m$. The coupling constant renormalization in (3.6), (3.7) is 
determined in the $\overline{\rm MS}$-scheme as far as the light flavours 
and the gluon are concerned. In addition we make the choice that the 
heavy quarks decouple in the running strong coupling constant 
$\alpha_s(\mu^2)$ for $\mu^2 < m_H^2$ and the renormalized OME's.
The factor $1 + \epsilon^2 \zeta(2)/8$ in (3.7) arises from the
requirement that $\Pi_H(0, m_H^2) = 0$, where $\Pi_H(p^2, m_H^2)$ is the
contribution to the gluon self-energy due to the heavy quark loops indicated
by $H$. 
After coupling constant renormalization $\hat A_{ij}$ takes
the form up to $O(\alpha_s^2)$
\begin{eqnarray}
\hat A_{ij} = \delta_{ij} + 
\Big( \frac{\alpha_s}{4 \pi}\Big) \hat A_{ij}^{(1)}
+ \Big( \frac{\alpha_s}{4 \pi}\Big)^2
\Big\{ \hat A_{ij}^{(2)} + \delta m \frac{d}{dm} \hat A_{ij}^{(1)} 
+ \delta \alpha_s \hat A_{ij}^{(1)} \Big\}\,.
\end{eqnarray}

The remaining UV-divergences are removed by operator renormalization which 
is achieved by
\begin{eqnarray}
\hat A_{ij} \Big( \epsilon_{\rm UV}, \epsilon_{\rm C},
\frac{\mu^2}{m^2}, \alpha_s \Big) =
Z_{ik}( \epsilon_{\rm UV}, \alpha_s)\otimes
\tilde A_{kj} \Big( \epsilon_{\rm C}, \frac{\mu^2}{m^2}, \alpha_s \Big) 
\,,
\end{eqnarray}
where $Z_{ij}$ $ (i,j = q,g)$ are the operator renormalization
constants corresponding to the operators in (2.24)-(2.26).
Notice that for the non-singlet operator in (2.24) $Z_{qq}^{\rm NS}$
is a real number, whereas for the singlet operators in (2.25), (2.26)
$Z_{ij}$ becomes a matrix. The operator matrix elements can be
expanded in $\alpha_s$ analogous to $A_{ij}$ in (3.1) as follows
\begin{eqnarray}
Z_{ij} = \sum_{k=0}^{\infty} \Big( \frac{\alpha_s}{4\pi} \Big)^{k}
Z_{ij}^{(k)} \,,
\end{eqnarray}
so that the renormalized OME's $\tilde A_{ij}$ in (3.11) read up
to $O(\alpha_s^2)$, (using $ Z_{ij}^{(0)} = \delta_{ij}$)
\begin{eqnarray}
&& \tilde A_{ij} = \delta_{ij} + 
\Big( \frac{ \alpha_s}{4 \pi}\Big) [\hat A_{ij}^{(1)} + Z_{ij}^{-1,(1)}]
+ \Big( \frac{ \alpha_s}{4 \pi}\Big)^2
\Big\{ \hat A_{ij}^{(2)} 
+ \delta m \frac{d}{dm} \hat A_{ij}^{(1)} 
\nonumber  \\ \cr && \qquad \qquad  
+ \delta\alpha_s \hat A_{ij}^{(1)}  + Z_{ik}^{-1,(1)} \hat A^{(1)}_{kj}
+ Z^{-1,(2)}_{ij} \Big\} \,,
\end{eqnarray}
where $Z_{ij}^{-1,(k)}$ are the expansion coefficients of the inverse
matrix $Z^{-1}$. Choosing the $\overline{\rm MS}$-scheme
one obtains the following expression (see \cite{mn}) up to $O(\alpha_s^2)$
\begin{eqnarray}
&& Z_{ij}( \epsilon_{\rm UV}, \alpha_s) = 
\delta_{ij}  
+ \Big( \frac{\alpha_s}{4 \pi}\Big) S_\epsilon
[ - \frac{1}{\epsilon_{\rm UV}} P^{(0)}_{ij}]
\nonumber \\ \cr && \qquad
+ \Big( \frac{\alpha_s}{4 \pi}\Big)^2 S_\epsilon^2 
[\frac{1}{\epsilon_{\rm UV}^2} \Big\{ \frac{1}{2} P^{(0)}_{ik}
\otimes P^{(0)}_{kj}
+( \beta_0 + \beta_{0,H} ) P^{(0)}_{ij} \Big\}
\nonumber \\ \cr && \qquad
 - \frac{1}{\epsilon_{\rm UV}} \delta\alpha_s P^{(0)}_{ij}
 - \frac{1}{2\epsilon_{\rm UV}} P^{(1)}_{ij}] \,.
\end{eqnarray}
Here $P^{(0)}_{ij}$ and $P^{(1)}_{ij}$ stand for the first and second order
AP-splitting functions \cite{glap} respectively 
which have been calculated in the literature (see \cite{glap}, \cite{frs},
\cite{gly}, \cite{fkl}, \cite{cfp}).
Notice that in our notation $A_{ij}$ and $Z_{ij}$ depend on the 
partonic Bjorken scaling variable $z$ in (2.8) 
which for convenience has been suppressed in
our formulae. Hence instead of multiplications we have to deal with
convolutions denoted by the symbol $\otimes$, which is defined in (2.21).
Furthermore the AP-splitting functions are related to the anomalous dimensions 
$\gamma^{(m)}_{ij}$ of the composite operators in (2.24)-(2.26)
via the Mellin transform
\begin{eqnarray}
\gamma^{(m),(k)}_{ij} = - \int^1_0 \, dz~z^{m-1} P^{(k)}_{ij}(z) \,.
\end{eqnarray}
Finally we have to remove the $C$-divergences appearing in
$\tilde A_{ij}$ (3.11). This is achieved by performing mass factorization
which proceeds as 
\begin{eqnarray}
\tilde A_{ij} \Big(  \epsilon_{\rm C}, \frac{\mu^2}{m^2}, \alpha_s \Big) =
A_{ik}\Big( \frac{\mu^2}{m^2}, \alpha_s \Big)\otimes
\Gamma_{kj} ( \epsilon_{\rm C}, \alpha_s ) \,,
\end{eqnarray}
where $A_{ij}$ denote the finite OME's which do not have UV or C
divergences anymore. The quantities $\Gamma_{ij}$ stand for the transition
functions which have the same properties as $Z_{ij}$ mentioned above (3.12).
Like $A_{ij}$ and $Z_{ij}$ the $\Gamma_{ij}$
can be expanded as a power series in $\alpha_s$  
\begin{eqnarray}
\Gamma_{ij} = \sum_{k=0}^{\infty} \Big( \frac{\alpha_s}{4\pi} \Big)^{k}
\Gamma_{ij}^{(k)} \,,
\end{eqnarray}
so that the mass factorized OME's $A_{ij}$ read up to $O(\alpha_s^2)$
(using $\Gamma^{(0)}_{ij} = \delta_{ij})$
\begin{eqnarray}
&& A_{ij} = \delta_{ij} + 
\Big( \frac{\alpha_s}{4 \pi}\Big) [\hat A_{ij}^{(1)} 
+ Z_{ij}^{-1,(1)} + \Gamma_{ij}^{-1,(1)}]
\nonumber \\ \cr && \qquad
+ \Big( \frac{ \alpha_s}{4 \pi}\Big)^2
\Big[  \hat A_{ij}^{(2)} + \delta m \frac{d}{dm} \hat A_{ij}^{(1)} 
+ \delta\alpha_s \hat A_{ij}^{(1)}  
+ Z_{ik}^{-1,(1)} \hat A_{kj}^{(1)} + Z_{ij}^{-1,(2)} 
\nonumber \\ \cr && \qquad
+ \Big\{ \hat A_{ik}^{(1)}  
+  Z_{ik}^{-1,(1)}\Big\} \otimes \Gamma^{-1,(1)}_{kj} 
+  \Gamma_{ij}^{-1,(2)} \Big] \,,
\end{eqnarray}
where $\Gamma^{-1,(k)}_{ij}$ are the expansion coefficients of the
inverse matrix $\Gamma^{-1}$. Choosing the
$\overline{\rm MS}$-scheme one obtains the following expression 
(see \cite{zn}) up to $O(\alpha_s^2)$
\begin{eqnarray}
&& \Gamma_{ij}( \epsilon_{\rm C}, \alpha_s) = 
\delta_{ij}  
+ \Big( \frac{\alpha_s}{4 \pi}\Big) S_\epsilon
[ \frac{1}{\epsilon_{\rm C}} P^{(0)}_{ij}]
\nonumber \\ \cr && \qquad
+ \Big( \frac{\alpha_s}{4 \pi}\Big)^2 S_\epsilon^2
[ \frac{1}{\epsilon_{\rm C}^2} \Big\{ \frac{1}{2} P^{(0)}_{ik}
\otimes P^{(0)}_{kj}
+ \beta_0 P^{(0)}_{ij} \Big\}
 + \frac{1}{2\epsilon_{\rm C}} P^{(1)}_{ij}] \,.
\end{eqnarray}
If all quarks would be massless we would have the identity
$\Gamma_{ij} = Z^{-1}_{ij}$.
However since the heavy quark $Q$ is massive it does not contribute to those 
splitting functions $P^{(k)}_{ij}$ which appear in the transition
functions. Hence in this case $i$ and $j$ only represent the light quarks and 
the gluon. This is in contrast to the operator renormalization constant 
$Z_{ij}$ where $i$ and $j$ can also stand for the heavy quark. 
The same assertion holds for
the heavy flavour contributions to the beta-function given by
$\beta_{0,H}$ ($m_H^2 \ge Q^2)$ which shows up in (3.14) but not in (3.19).
The reason is that the mass $m$ of the heavy quark acts as a regulator
for the C-divergences but not for the UV-singularities. 

Using the master formula in (3.18) one can now construct
the general form of the finite OME $A_{ij}$  as well as the 
unrenormalized OME $\hat A_{ij}$  
expressed in their renormalization group
coefficients $\beta_0$, $\beta_{0,H}$ 
and $P^{(k)}_{ij}$ ($k=0,1$; $ i,j = q,g$).
For convenience we will take for $\hat A_{ij}$
the representation (3.3), where the mass renormalization has been 
already carried out.

Starting with the one-loop OME $\hat A^{(1)}_{Qg}$, which receives
contributions from the graphs in fig.1, we obtain
\begin{eqnarray}
\hat  A^{(1)}_{Qg}  = S_\epsilon \Big(\frac{m^2}{\mu^2}\Big)^{\epsilon/2}
\Big\{ - \frac{1}{\epsilon_{\rm UV}} P^{(0)}_{qg} + a^{(1)}_{Qg} +
\epsilon_{\rm UV} \bar a^{(1)}_{Qg} \Big\} \,.
\end{eqnarray}
The renormalization group coefficients are given by
\begin{eqnarray}
&& P^{(0)}_{qg} = 8 T_f [ z^2 + (1-z)^2 ] \, ,
\nonumber  \\ \cr && 
a^{(1)}_{Qg} = 0 \, ,
\nonumber  \\ \cr && 
\bar a^{(1)}_{Qg} = - \frac{1}{8} \zeta(2) P^{(0)}_{qg}\,.
\end{eqnarray}
The expression for the two-loop contribution to $\hat A_{Qg}$ (see fig.2)
can be written as
\begin{eqnarray}
&& \hat A_{Qg}^{(2)}  = S_\epsilon^2 \Big( \frac{m^2}{\mu^2}\Big)^\epsilon
\Big[ \frac{1}{\epsilon^2} \Big\{
\frac{1}{2} P^{(0)}_{qg} \otimes ( P^{(0)}_{qq} - P^{(0)}_{gg})
+ \beta_0 P^{(0)}_{qg} \Big\}
\nonumber  \\ \cr && \qquad 
+ \frac{1}{\epsilon} \Big\{ - \frac{1}{2} P^{(1)}_{qg}
-2 \beta_0 a^{(1)}_{Qg} - a^{(1)}_{Qg} \otimes (P^{(0)}_{qq} - P^{(0)}_{gg})
\Big\} + a^{(2)}_{Qg} \Big]
\nonumber  \\ \cr && \qquad
- \frac{2}{\epsilon} S_\epsilon \beta_{0,H}
\Big( \frac{m_H^2}{\mu^2} \Big)^{\epsilon/2}
\Big( 1 + \frac{\epsilon^2}{8} \zeta(2) \Big)
\hat A^{(1)}_{Qg} \, .
\end{eqnarray}
Notice that in the above expression the pole terms $\epsilon^{-k}$
stand for the UV as well as C-divergences so that we have
put $\epsilon_{\rm UV} = \epsilon_{\rm C}$. Further we infer from the
literature that
\begin{eqnarray}
&&P^{(0)}_{qq}=4 C_F [ 2 \Big(\frac{1}{1-z}\Big)_+
-1-z+\frac{3}{2}\delta(1-z)]\,,
\nonumber  \\ \cr && 
P^{(0)}_{gg} = 8 C_A [ \Big(\frac{1}{1-z}\Big)_+ 
+ \frac{1}{z} - 2 + z - z^2 ] + 2 \beta_0 \delta(1-z)\,,
\nonumber  \\ \cr && 
P^{(1)}_{qg} = 8 C_F T_f [2(1-2z+2z^2)\{\ln^2(1-z) - 2 \ln z \ln (1-z)
- 2 \zeta(2)\}
\nonumber  \\ \cr && \qquad
+(1-2z+4z^2) \ln^2 z + 8z (1-z) \ln (1-z) + (3 -4z +8z^2)\ln z
\nonumber  \\ \cr && \qquad
+ 14 - 29 z + 20 z^2]
\nonumber  \\ \cr &&  \qquad 
+ 8 C_A T_f \Big[(1+2z+2z^2)\{\ln^2 z - 4 \ln z \ln (1+z) - 4 {\rm Li}_2(-z)
- 2 \zeta(2)\}
\nonumber  \\ \cr && \qquad
+2(1-2z+2z^2)[\zeta(2) - \ln^2(1-z)]
-(3+6z+2z^2)\ln^2 z 
\nonumber  \\ \cr && \qquad
- 8z (1-z) \ln(1-z) 
+\Big(2 + 16z +\frac{88}{3}z^2\Big) \ln z  
\nonumber  \\ \cr && \qquad
+ \frac{2}{9} \Big(\frac{20}{z} -  18 + 225 z - 218 z^2 \Big)\Big]  \,,
\end{eqnarray}
and $a^{(2)}_{Qg} $ (3.22) has to be computed in this paper. In the above
and subsequent expressions the functions ${\rm Li}_n(z)$ denote the 
polylogarithms which can be found in \cite{lbmr}. The finite
OME's follow from (3.18) where we have
\begin{eqnarray}
&& A^{(1)}_{Qg} = \hat A^{(1)}_{Qg} + Z^{-1,(1)}_{qg}
\nonumber  \\ \cr && \qquad  
= - \frac{1}{2} P^{(0)}_{qg} \ln \frac{m^2}{\mu^2} 
+ a^{(1)}_{Qg} \,,
\end{eqnarray}
and
\begin{eqnarray}
&& A_{Qg}^{(2)} =  \hat A_{Qg}^{(2)} + \delta\alpha_s \hat A_{Qg}^{(1)}  
+ Z^{-1,(1)}_{qq} \otimes  \hat A_{Qg}^{(1)}  
+ Z^{-1,(1)}_{qg} \otimes  \hat A_{gg}^{(1)}  
\nonumber  \\ \cr && \qquad 
+  Z_{qg}^{-1,(2)} +  ( \hat A_{Qg}^{(1)} +  Z^{-1,(1)}_{qg})
\otimes \Gamma^{-1,(1)}_{gg}
\nonumber  \\ \cr && \qquad
= \Big\{ 
  \frac{1}{8}   P^{(0)}_{qg} \otimes P^{(0)}_{qq}
- \frac{1}{8}   P^{(0)}_{qg} \otimes P^{(0)}_{gg}
 + \frac{1}{4}  \beta_0 P^{(0)}_{qg} \Big\} \ln^2 \frac{m^2}{\mu^2}
\nonumber  \\ \cr && \qquad
+ \Big\{ 
- \frac{1}{2}   P^{(1)}_{qg}  - \beta_0 a^{(1)}_{Qg} 
- \frac{1}{2}   P^{(0)}_{qq} \otimes a^{(1)}_{Qg}
 + \frac{1}{2}  P^{(0)}_{gg} \otimes a^{(1)}_{Qg}\Big\} \ln \frac{m^2}{\mu^2}
\nonumber  \\ \cr && \qquad
+a^{(2)}_{Qg} + 2 \beta_0 \bar a^{(1)}_{Qg} + P^{(0)}_{qq} \otimes
\bar a^{(1)}_{Qg} - P^{(0)}_{gg} \otimes \bar a^{(1)}_{Qg} \,.
\end{eqnarray}
In eqs.(3.24), (3.25) we have used the heavy flavour contributions 
($m_H^2 \ge Q^2$) to the one loop 
OME $A_{gg}^{(1)}$ in fig.1c which is given by
\begin{eqnarray}
 A_{gg}^{(1)}  = 
S_\epsilon \Big( \frac{m_H^2}{\mu^2} \Big)^{\epsilon/2}
\Big[-\frac{2}{\epsilon_{\rm UV}} \beta_{0,H} - \frac{1}{4} 
\epsilon_{\rm UV} \beta_{0,H} \zeta(2) \Big] \,,
\end{eqnarray}
and the constants
\begin{eqnarray}
Z_{qg}^{-1, (1)}  = 
S_\epsilon \Big[  \frac{1}{\epsilon_{\rm UV}} P^{(0)}_{qg} \Big] \,,
\nonumber 
\end{eqnarray}
\begin{eqnarray}
&& Z_{qg}^{-1,(2)}  = 
S_\epsilon^2  \Big[ \frac{1}{\epsilon_{\rm UV}^2} \Big\{
\frac{1}{2} P^{(0)}_{qg} \otimes ( P^{(0)}_{qq} +
P^{(0)}_{gg}) - \beta_0  P^{(0)}_{qg}\Big\}
\nonumber  \\ \cr && \qquad
+ \frac{1}{\epsilon_{\rm UV}} 
\delta\alpha_s P^{(0)}_{qg} + \frac{1}{2\epsilon_{\rm UV}} 
 P^{(1)}_{qg}  \Big] \,,
\nonumber 
\end{eqnarray}
\begin{eqnarray}
Z_{qq}^{-1, (1)}  = 
S_\epsilon \Big[  \frac{1}{\epsilon_{\rm UV}} P^{(0)}_{qq} \Big] \,,
\nonumber 
\end{eqnarray}
\begin{eqnarray}
\Gamma_{gg}^{-1, (1)}  = 
S_\epsilon \Big[ -\frac{1}{\epsilon_{\rm C}} P^{(0)}_{gg} \Big] \,.
\end{eqnarray}
Notice that due to our scheme for $\alpha_s$ in (3.7) $\beta_{0,H}$ and $m_H$
$(m_H^2 \ge Q^2)$ have completely disappeared from $A^{(2)}_{Qg}$ (3.25).

The renormalization and the mass factorization of the OME's
$\hat A^{\rm PS,(2)}_{Qq}$ in fig.3 proceeds in the same way as done for
$\hat A_{Qg}$.
The superscript PS stands for 'pure singlet' and it originates
from the definition that the singlet OME $A^{\rm S}_{qq}$
can be decomposed into
\begin{eqnarray}
 A_{qq}^{\rm S}  = A_{qq}^{\rm NS}  + A_{qq}^{\rm PS}   \,,
\end{eqnarray}
where $A_{qq}^{\rm NS}$ stands for the non-singlet OME. The unrenormalized
OME reads
\begin{eqnarray}
&& \hat A_{Qq}^{{\rm PS},(2)}  = 
S_\epsilon^2 \Big( \frac{m^2}{\mu^2} \Big)^{\epsilon}
\Big[ \frac{1}{\epsilon^2} \Big\{-\frac{1}{2} 
P^{(0)}_{qg}\otimes P^{(0)}_{gq} \Big\}
+ \frac{1}{\epsilon} \Big\{
-\frac{1}{2} P^{{\rm PS},(1)}_{qq}
\nonumber  \\ \cr && \qquad \qquad
+ a^{(1)}_{Qg}\otimes P^{(0)}_{gq} \Big\}
+ a^{{\rm PS}, (2)}_{Qq} \Big] \,.
\end{eqnarray}
Like in the case of $\hat A_{Qg}^{(2)}$  we did not make any distinction 
between UV and C-singular pole terms $\epsilon^{-k}$ $(\epsilon_{\rm UV}=
\epsilon_{\rm C})$. The renormalization group coefficients are given by
(see also (3.21))
\begin{eqnarray}
&& P^{(0)}_{gq} = 4 C_F [ \frac{2}{z}  -2 + z  ] \,,
\nonumber  \\ \cr &&   
 P^{{\rm PS},(1)}_{qq} = 8 C_F T_f \Big[ -2(1+  z)\ln^2 z
+ \Big(2 + 10 z + \frac{16}{3} z^2 \Big) \ln z  
\nonumber  \\ \cr && \qquad \qquad 
+ \frac{40}{9z} - 4 + 12 z - \frac{112}{9} z^2 \Big] \,,
\end{eqnarray}
and $a^{\rm PS, (2)}_{Qq}$ 
will be calculated in this paper. The finite OME $A^{\rm PS, (2)}_{Qq}$ 
can be again derived from (3.18) which yields
\begin{eqnarray}
&& A_{Qq}^{{\rm PS},(2)} = \hat A_{Qq}^{{\rm PS},(2)} 
+\Big(Z^{\rm PS}_{qq}\Big)^{-1,(2)}  +  
\Big( \hat A_{Qg}^{(1)} + Z_{qg}^{-1,(1)}\Big)
\otimes \Gamma^{-1,(1)}_{gq}
\nonumber \\ \cr && \qquad
=  \Big\{ - \frac{1}{8} P^{(0)}_{qg} \otimes P^{(0)}_{gq} \Big\} \ln^2 \frac
{m^2}{\mu^2}
\nonumber \\ \cr && \qquad
+   \Big\{ - \frac{1}{2} P^{{\rm PS},(1)}_{qq} 
+ \frac{1}{2} a^{(1)}_{Qg}\otimes P^{(0)}_{gq} \Big\} \ln \frac {m^2}{\mu^2}
\nonumber \\ \cr && \qquad
+ a^{{\rm PS}, (2)}_{Qq} - \bar a^{(1)}_{Qg} \otimes P^{(0)}_{gq} ]\,.
\end{eqnarray}
Here we have used the renormalization group constants (see also (3.21),(3.30))
\begin{eqnarray}
\Big(Z^{\rm PS}_{qq}\Big)^{-1,(2)} = S_\epsilon^2 \Big[ \frac{1}
{\epsilon_{\rm UV}^2}
\Big\{ \frac{1}{2} P^{(0)}_{qg} \otimes P^{(0)}_{gq} \Big\}
+ \frac{1}{2\epsilon_{\rm UV}} P^{{\rm PS},(1)}_{qq} \Big]\,, 
\nonumber
\end{eqnarray}
\begin{eqnarray}
\Gamma_{gq}^{-1,(1)} = S_\epsilon \Big[-\frac{1}{\epsilon_{\rm C}}
P^{(0)}_{gq} \Big]\,. 
\end{eqnarray}

The last OME which we have to deal with is represented by
the heavy quark loop contribution to $A^{\rm NS, (2)}_{qq}$ 
in fig.4. The unrenormalized expression reads
\begin{eqnarray}
\hat A_{qq,Q}^{{\rm NS},(2)}=S_\epsilon^2 \Big( \frac{m^2}{\mu^2} 
\Big)^\epsilon \Big[ \frac{1}{\epsilon_{\rm UV}^2}
\{ - \beta_{0,Q} P^{(0)}_{qq} \}
+ \frac{1}{\epsilon_{\rm UV}}
\{ -\frac{1}{2} P^{{\rm NS},(1)}_{qq,Q}\}
+a^{{\rm NS},(2)}_{qq,Q}\Big] \,.
\end{eqnarray}
Contrary to $\hat A^{(2)}_{Qg}$ and $A^{\rm PS, (2)}_{Qq}$ 
the above OME only contains UV-divergences since the heavy quark $Q$
prevents $\hat A^{{\rm NS},(2)}_{qq,Q}$  to be C-singular provided we choose 
the coupling constant renormalization scheme in (3.7).
The renormalization group coefficients in (3.33) are given by
\begin{eqnarray}
\beta_{0,Q} =  - \frac{4}{3} T_f \, ,
\nonumber 
\end{eqnarray}
\begin{eqnarray}
&& P^{{\rm NS},(1)}_{qq,Q}=C_F T_f\Big[ - \frac{160}{9}
\Big( \frac{1}{1-z}\Big)_+
+ \frac{176}{9} z - \frac{16}{9} - \frac{16}{3} \frac{1+z^2}{1-z} \ln z
\nonumber  \\ \cr && \qquad \qquad 
+ \delta(1-z) 
\Big( - \frac{4}{3} - \frac{32}{3} \zeta(2)\Big) \Big] \, .
\end{eqnarray}
Since $\hat A^{{\rm NS},(2)}_{qq}$ does not contain C-divergences we only
need operator renormalization to render it finite. From (3.18) we infer
\begin{eqnarray}
&& A^{{\rm NS},(2)}_{qq,Q}  =  \hat A^{{\rm NS},(2)}_{qq}  +
\Big( Z^{\rm NS}_{qq,Q} \Big)^{-1,(2)} 
\nonumber  \\ \cr && \qquad
= \Big\{ - \frac{1}{4} \beta_{0,Q} P^{(0)}_{qq} \Big\} \ln^2 \frac{m^2}{\mu^2}
+ \Big\{ - \frac{1}{2} P^{{\rm NS},(1)}_{qq,Q}\Big\} \ln\frac{m^2}{\mu^2}
\nonumber  \\ \cr && \qquad
+ a^{{\rm NS},(2)}_{qq,Q} + \frac{1}{4} \beta_{0,Q} \zeta(2) P^{(0)}_{qq}\,,
\end{eqnarray}
with
\begin{eqnarray}
&& \Big(Z^{\rm NS}_{qq,Q}\Big)^{-1,(2)}  = S_\epsilon^2 [
 - \frac{1}{\epsilon_{\rm UV}^2}
 \beta_{0,Q} P^{(0)}_{qq} + \frac{2}{\epsilon_{\rm UV}^2} \beta_{0,Q}
\Big( \frac{m^2}{\mu^2}\Big)^{\epsilon/2}
\Big( 1 + \frac{1}{8} \zeta(2) \epsilon_{\rm UV}^2 \Big) P^{(0)}_{qq}
\nonumber  \\ \cr && \qquad \qquad \qquad
+ \frac{1}{2\epsilon_{\rm UV}} P^{{\rm NS},(1)}_{qq,Q} ]\,.
\end{eqnarray}
Notice that expressions (3.33) and (3.35) also contribute via (3.28) to the 
singlet part of the OME.

Since we can infer the coefficients of the double and single pole terms of the
unrenormalized OME $\hat A_{ij}$ from the two-loop corrected AP-splitting 
functions and the beta-function,
the above expressions serve as a check on our calculations.
The non-pole terms defined by $a^{(2)}_{ij}$, which cannot be
predicted, have to be calculated in this paper.
They are needed to compute the heavy quark coefficient functions
(2.18) up to the non-logarithmic term, which will be done in the 
next section.

Before finishing this section we will give an outline of our calculation
of the OME's depicted in figs.1-4. 
We have computed the OME using the standard
QCD Feynman rules and the operator vertices
corresponding to (2.24), (2.25) which can be found in the literature
(see \cite{frs}). 
Since in our case the latter are S-matrix elements we have to consider
the connected Green's functions where the external quark and gluon propagators
are amputated. Notice that one has to include the external self
energies. The connected Green's function needed for
$A_{qg}$ (figs.1,2) is given by
\begin{eqnarray} 
< 0 | T( A^a_\mu (x) O^{\mu_1 ... \mu_m}_{q} (0) A^b_\nu(y) |0>_c \,,
\end{eqnarray}
and for $A^{\rm PS}_{qq}$ (fig.3) and
$A^{\rm NS}_{qq}$ (fig.4) we have 
\begin{eqnarray} 
< 0 | T( \bar\psi_i(x) O^{r,\mu_1 ... \mu_m}_{q} (0) \psi_j(y) |0>_c \,,
\end{eqnarray}
with $r=$  PS and $r=$ NS respectively. Further $a,b$ and $i,j$ are the colour
indices of the gluon field $A_\mu$ and the quark field $\psi$
respectively.
Since the above operators $O^{\mu_1....\mu_m}(0)$ are traceless
symmetric tensors under the Lorentz group the computation of the connected
Green's functions reveals the presence of many trace
terms
which are not essential for the determination of the anomalous
dimensions or splitting functions. Therefore  we will project them out by
multiplying the operators by an external source
$J_{\mu_1....\mu_m}  = \Delta_{\mu_1}...\Delta_{\mu_m}$ with $\Delta^2 = 0$.
Performing the Fourier transform into momentum space and sandwiching
the connected Green's functions by the external gluon
polarizations and quark spinors one obtains
\begin{eqnarray} 
\epsilon^\mu(p) G^{ab}_{q,\mu\nu} \epsilon^\nu(p) \,,
\end{eqnarray}
and
\begin{eqnarray} 
 \bar u(p,s) G^{ij}_{q} \lambda_r u(p,s) \,,
\end{eqnarray}
where $\lambda_r$ denote the generators of the flavour group SU$(n_f)$.
The tensors $G^{ab}_{q,\mu\nu}$ and $G^{ij}_q$ have the form
\begin{eqnarray} 
G^{ab}_{q,\mu\nu} = \hat A^{(m)}_{qg}\Big(\epsilon, \frac{m^2}{\mu^2},
\alpha_s\Big) 
\delta^{ab}(\Delta\cdot p)^m \Big( - g_{\mu\nu}
+ \frac{\Delta_\mu p_\nu + \Delta_\nu p_\mu}{\Delta\cdot p} \Big) \,,
\end{eqnarray}
and 
\begin{eqnarray} 
G^{ij}_{q} = \hat A_{qq} \Big( \epsilon, \frac{m^2}{\mu^2} , \alpha_s\Big)
 \delta^{ij} ( \Delta\cdot p)^{m-1} {\Delta \llap /} \,.
\end{eqnarray}
By projecting these tensors out one obtains finally the OME's
\begin{eqnarray} 
\hat A^{(m)}_{qg}\Big(\epsilon, \frac{m^2}{\mu^2}, \alpha_s\Big) 
=  \frac{1}{N^2-1} \frac{1}{n-2} (-g_{\mu\nu}) \delta^{ab} 
(\Delta\cdot p)^{-m} G^{ab}_{q,\mu\nu} \, , 
\end{eqnarray}
\begin{eqnarray} 
\hat A_{qq}^{(m)} \Big( \epsilon, \frac{m^2}{\mu^2} , \alpha_s\Big)
= \frac{1}{N} \delta^{ij} \frac{1}{4} ( \Delta\cdot p)^{-m} {\rm Tr}
({p\llap /}  G^{ij}_q) \,.
\end{eqnarray}
Notice that in (3.43) the summation over the dummy indices
$\mu$ and $\nu$ includes unphysical (non-transverse) gluon polarizations
which have to be compensated by adding graphs containing 
the external ghost lines
represented in fig.2s and fig.2t. Instead of $-g^{\mu\nu}/(n-2)$
one can also use the physical polarization sum
$[-g_{\mu\nu} + ( \Delta_\mu p_\nu + \Delta_\nu p_\mu)/\Delta \cdot p]/(n-2)$
and omit the graphs with the external ghost lines. However in this case the
individual Feynman graphs lead to integrals with higher powers
of the momenta in the numerator so that they become
more difficult to compute. Moreover the number of independent
integrals is artificially increased. The advantage of constructing
the Green's function in the way shown in (3.43), (3.44) is that one
does not have to resort to complicated tensorial reduction
programs as had to be used for example in \cite{mn}. Therefore one has only
Lorentz scalars in the numerators of the Feynman integrals which can be 
partially cancelled by terms in the denominators.
The traces of the fermion loops in figs.1,2 and the contraction with the 
metric tensor in (3.43) have been performed by using
the algebraic manipulation program FORM \cite{jamv}.
We did the same for the graphs in figs.3,4 where we had to compute
the trace in (3.44).
The computation of the scalar integrals is straightforward as long as the
number of propagators does not exceed four. 
In the case of five different propagators the calculation
becomes more cumbersome but here one can use the trick
of integration by parts \cite{ct}. Examples are given in Appendix B.
The results for the unrenormalized OME's are too long to put them
in this section and we will defer them to Appendix C.

Apart from the check on the pole terms mentioned above we can also
check the finite term of the Abelian part of $\hat A_{Qg}$. Removing
the overall constant $(1 + (-1)^m)/2$ the Mellin transform reads up to order
$\alpha_s^2$ (see Appendix C)
\begin{eqnarray}
&& \hat A^{(m)}_{Qg} = 
S_\epsilon \Big( \frac{\alpha_s}{4\pi}\Big)
T_f \Big[ \hat A^{(1),(m)}_{Qg}\Big]
+ S_\epsilon^2 \Big( \frac{\alpha_s}{4\pi}\Big)^2 T_f \Big[ C_F 
\hat A^{(2),(m)}_{Qg,F} + C_A
\hat A^{(2),(m)}_{Qg,A}\Big]  \,.
\nonumber \\ &&
\end{eqnarray}
Notice that coupling constant renormalization has already been carried out 
in the above equation 
so that we get rid of the term proportional
to $T_f^2$ in (C.1). If we now take the first moment
$m=1$, $\hat A^{(2),(m)}_{Qg,A} \approx (m-1)^{-1}$ but
$\hat A^{(1),(1)}_{Qg}  $ and $\hat A^{(2),(1)}_{Qg,F}$ are finite.
One can now easily show that there exists a relation between the Abelian
terms and the heavy fermion loop contribution to the gluon self energy
which we will denote by 
\begin{eqnarray} 
\hat \Pi_V(p^2,m^2) = 
S_\epsilon \Big( \frac{\alpha_s}{4\pi}\Big)
T_f \hat \Pi_V^{(1)}(p^2,m^2)  
+ S_\epsilon^2 \Big( \frac{\alpha_s}{4\pi}\Big)^2
C_F T_f \hat \Pi_V^{(2)}(p^2,m^2) \,.
\end{eqnarray}
The relation is given by
\begin{eqnarray} 
\hat \Pi_V^{(1)}(0,m^2) =  \frac{1}{2} \hat A^{(1),(1)}_{Qg}\,,
\end{eqnarray}
\begin{eqnarray} 
\hat \Pi_V^{(2)}(0,m^2) =  \frac{1}{2} \hat A^{(2),(1)}_{Qg,F}\,.
\end{eqnarray}
$\hat \Pi_V^{(1)}(0,m^2) $ and $\hat \Pi_V^{(2)}(0,m^2) $ can be inferred from
eqs. (2.11) and (5.1) in \cite{dg}. In the latter reference 
the heavy fermion loop contribution to the self energy of the
$Z$-boson and the photon were calculated. 
The self-energy contribution to the gluon
and the photon are related as follows
\begin{eqnarray} 
\hat \Pi_V^{(1)}(p^2,m^2) =   4 \frac{1}{p^2} \Pi^V_T(p^2)\,,
\end{eqnarray}
\begin{eqnarray} 
\hat \Pi_V^{(2)}(p^2,m^2) =   4^2 \frac{3}{4} \frac{1}{3} 
\frac{1}{p^2} \Pi^V_T(p^2)\,,
\end{eqnarray}
where $\Pi^V_T(p^2)$ in eqs.(3.49), (3.50) are equal to the expressions
quoted in (2.11) and (5.1) of \cite{dg} respectively.
The factors of 4 originate from the convention that in \cite{dg}
one expands the quantities in $\alpha_s/\pi$ instead of $\alpha_s/(4\pi)$
as is done in this paper. The factors of 4/3 and 3 refer to the colour 
factors $C_F$ and $C_A$ in SU(3). From \cite{dg} we infer that
\begin{eqnarray} 
\hat \Pi_V^{(1)}(0,m^2) =   \Big( \frac{m^2}{\mu^2}\Big)^{\epsilon/2}
\Big[ - \frac{8}{3\epsilon} \Big] \,,
\end{eqnarray}
\begin{eqnarray} 
\hat \Pi_V^{(2)}(0,m^2) =   \Big( \frac{m^2}{\mu^2}\Big)^{\epsilon/2}
\Big[ - \frac{4}{\epsilon}  + 15 \Big] \,,
\end{eqnarray} 
where $n=4 + \epsilon$. (Notice that in \cite{dg} $n= 4 - 2\epsilon$).
Calculation of the first moment of (3.45) (see also (C.1),(C.2)) 
reveals that the relations in (3.47), (3.48) are satisfied.
\vfill
\newpage
\mysection{Heavy quark coefficient functions}
In this section we will compute the heavy quark coefficient functions
$H_{i,l}$ and $L_{i,l}$ defined above (2.14)
up to order $\alpha_s^2$ in the asymptotic limit $Q^2 \gg  m^2$.
For this computation we will use the mass factorization theorems
as represented by eqs. (2.19) and (2.20) where the
transition functions $\Gamma_{kl}(\mu^2/m^2)$ stand for the finite
OME's $A_{kl}(\mu^2/m^2)$ computed in section 3.

Let us start with the heavy quark coefficient functions which
originate from the subprocesses where the virtual photon couples to
the heavy quark. Here the mass factorization
theorem implies
\begin{eqnarray}
H_{i,l}\Big(\frac{Q^2}{m^2}, \frac{\mu^2}{m^2}\Big)
= A_{kl}\Big(\frac{\mu^2}{m^2}\Big) \otimes C_{i,k}
\Big(\frac{Q^2}{\mu^2}\Big)\,,
\end{eqnarray}
where $H_{i,l}$ $(i=2,L; l=q,g)$ denote the heavy quark coefficient
functions in the limit $Q^2 \gg  m^2$ and $A_{kl}$ $(k,l = q,g)$
are the finite OME's computed in the last section. The coefficient
functions $C_{i,l}(Q^2/\mu^2)$ have been calculated in
\cite{zn}. They are obtained from the massless parton structure functions
as defined in eqs. (2.27), (2.28), after having performed mass 
factorization in the $\overline{\rm MS}$-scheme. In the last section
also the OME's $A_{kl}$ have been calculated in the
$\overline{\rm MS}$-scheme. In the product on the right-hand-side of (4.1)
the scheme dependence is only partially cancelled which is revealed by the fact
that $H_{i,l}$ is still scheme dependent. This dependence, which
is indicated by $\mu^2/m^2$ in $H_{i,l}\,,$ originates from the coupling 
of a light parton (gluon or quark) to an internal light parton 
characteristic of the production mechanisms of the processes in
(2.12) and (2.13).

In the case of the lowest order photon-gluon fusion process we have 
(see (2.11) and (2.14))
\begin{eqnarray}
H^{(1)}_{L,g}\Big(\frac{Q^2}{m^2},\frac{\mu^2}{m^2}\Big) 
= C^{(1)}_{L,g}\Big(\frac{Q^2}{\mu^2}\Big)\,,
\end{eqnarray}
\begin{eqnarray}
H^{(1)}_{2,g} \Big(\frac{Q^2}{m^2},\frac{\mu^2}{m^2}\Big) 
= A^{(1)}_{Qg}\Big(\frac{\mu^2}{m^2}\Big) 
+ C^{(1)}_{2,g}\Big(\frac{Q^2}{\mu^2}\Big) \,,
\end{eqnarray}
where $C^{(1)}_{L,g}$, $C^{(1)}_{2,g}$ are the order
$\alpha_s$ gluonic coefficient functions which e.g. can be found in \cite{zn}.
Like the OME's they can be expressed in the renormalization group 
coefficients as
\begin{eqnarray}
C^{(1)}_{L,g}\Big(\frac{Q^2}{\mu^2}\Big) = c^{(1)}_{L,g} \, ,
\end{eqnarray}
\begin{eqnarray}
C^{(1)}_{2,g} \Big(\frac{Q^2}{\mu^2}\Big) = 
\frac{1}{2} P^{(0)}_{qg}\ln \frac{Q^2}{\mu^2} + c^{(1)}_{2,g} \,.
\end{eqnarray}
From (3.24) and (4.2)-(4.5) one obtains the asymptotic
forms for the order $\alpha_s$ heavy quark coefficient
functions
\begin{eqnarray}
H^{(1)}_{L,g}\Big(\frac{Q^2}{m^2},\frac{\mu^2}{m^2}\Big) = c^{(1)}_{L,g} \,,
\end{eqnarray}
\begin{eqnarray}
H^{(1)}_{2,g} \Big(\frac{Q^2}{m^2},\frac{\mu^2}{m^2}\Big) = 
\frac{1}{2} P^{(0)}_{qg}\ln\frac{Q^2}{m^2} + a^{(1)}_{Qg} 
+ c^{(1)}_{2,g} \,.
\end{eqnarray}
In order $\alpha_s^2$ the coefficients of the 
photon-gluon fusion process (2.12) become
\begin{eqnarray}
H^{(2)}_{L,g}\Big(\frac{Q^2}{m^2}, \frac{\mu^2}{m^2}\Big) = A^{(1)}_{Qg}
\Big(\frac{\mu^2}{m^2}\Big) \otimes C^{(1)}_{L,q}\Big(\frac{Q^2}{\mu^2}\Big)
+ C^{(2)}_{L,g}\Big(\frac{Q^2}{\mu^2}\Big) \,,
\end{eqnarray}
\begin{eqnarray}
H^{(2)}_{2,g} \Big(\frac{Q^2}{m^2},\frac{\mu^2}{m^2}\Big) = 
A^{(2)}_{Qg}\Big(\frac{\mu^2}{m^2}\Big)+A^{(1)}_{Qg}\Big(\frac{\mu^2}{m^2}\Big) 
\otimes C^{(1)}_{2,q}\Big(\frac{Q^2}{\mu^2}\Big) 
+ C^{(2)}_{2,g}\Big(\frac{Q^2}{\mu^2}\Big)\,,
\end{eqnarray}
where $C^{(1)}_{i,q}$, $C^{(2)}_{i,g}$, $(i=2,L)$ become (see \cite{zn})
\begin{eqnarray}
C^{(1)}_{L,q}\Big(\frac{Q^2}{\mu^2}\Big) = c^{(1)}_{L,q} \, ,
\end{eqnarray}
\begin{eqnarray}
C^{(1)}_{2,q} \Big(\frac{Q^2}{\mu^2}\Big) = 
\frac{1}{2} P^{(0)}_{qq}\ln\frac{Q^2}{\mu^2} + c^{(1)}_{2,q} \, ,
\end{eqnarray}
\begin{eqnarray}
C^{(2)}_{L,g}\Big(\frac{Q^2}{\mu^2}\Big)  = \Big\{ - \beta_0 c^{(1)}_{L,g} 
+ \frac{1}{2} P^{(0)}_{gg}\otimes c^{(1)}_{L,g}
+ \frac{1}{2} P^{(0)}_{qg}\otimes c^{(1)}_{L,q} \Big\}
\ln\frac{Q^2}{\mu^2}
+  c^{(2)}_{L,g}\,,
\nonumber  \\  
\end{eqnarray}
\begin{eqnarray}
&& C^{(2)}_{2,g}\Big(\frac{Q^2}{\mu^2}\Big) = 
\Big\{ \frac{1}{8} P^{(0)}_{qg}\otimes( P^{(0)}_{gg} + P^{(0)}_{qq})
- \frac{1}{4} \beta_0 P^{(0)}_{qg}\Big\} \ln^2 \frac{Q^2}{\mu^2}
\nonumber \cr && \qquad \qquad
+ \Big\{  \frac{1}{2} P^{(1)}_{qg} - \beta_0 c^{(1)}_{2,g}
+ \frac{1}{2} P^{(0)}_{gg}\otimes c^{(1)}_{2,g}
+ \frac{1}{2} P^{(0)}_{qg}\otimes c^{(1)}_{2,q} \Big\}
\ln\frac{Q^2}{\mu^2}
\nonumber \\  && \qquad \qquad
+ c^{(2)}_{2,g}  \,.
\end{eqnarray}
From (3.24), (3.25) and (4.10)-(4.13) one infers the asymptotic
form
\begin{eqnarray}
&& H^{(2)}_{L,g}\Big(\frac{Q^2}{m^2},\frac{\mu^2}{m^2}\Big) = 
\Big\{ - \beta_0 c^{(1)}_{L,g} 
+ \frac{1}{2} P^{(0)}_{gg} \otimes c^{(1)}_{L,g} 
+ \frac{1}{2} P^{(0)}_{qg} \otimes c^{(1)}_{L,q} 
\Big\} \ln \frac{Q^2}{m^2}
\nonumber \cr && \qquad\qquad
+ \Big\{   - \beta_0 c^{(1)}_{L,g}
+ \frac{1}{2} P^{(0)}_{gg}\otimes c^{(1)}_{L,g} \Big\} \ln\frac{m^2}{\mu^2}
+ c^{(2)}_{L,g}  + a^{(1)}_{Qg} \otimes c^{(1)}_{L,q} \,, 
\nonumber \\ 
\end{eqnarray}
\begin{eqnarray}
&& H^{(2)}_{2,g}\Big(\frac{Q^2}{m^2}, \frac{\mu^2}{m^2}\Big) = 
\Big\{ \frac{1}{8} P^{(0)}_{qg}\otimes( P^{(0)}_{gg} + P^{(0)}_{qq})
- \frac{1}{4} \beta_0 P^{(0)}_{qg}\Big\} \ln^2 \frac{Q^2}{m^2}
\nonumber \cr && \qquad
+ \Big\{  \frac{1}{2} P^{(1)}_{qg} - \beta_0 c^{(1)}_{2,g}
+ \frac{1}{2} P^{(0)}_{qq}\otimes a^{(1)}_{Qg}
+ \frac{1}{2} P^{(0)}_{gg}\otimes c^{(1)}_{2,g} 
+ \frac{1}{2} P^{(0)}_{qg}\otimes c^{(1)}_{2,q} 
\Big\} 
\nonumber \cr && \qquad 
\times \ln\frac{Q^2}{m^2}
+\Big\{ \frac{1}{4} P^{(0)}_{qg} \otimes P^{(0)}_{gg} 
- \frac{1}{2} \beta_0 P^{(0)}_{qg}\Big\} \ln\frac{Q^2}{m^2} 
\ln \frac{m^2}{\mu^2}
\nonumber \cr && \qquad 
+\Big\{ -\beta_0( c^{(1)}_{2,g} + a^{(1)}_{Qg})
+ \frac{1}{2} P^{(0)}_{gg} \otimes ( c^{(1)}_{2,g} + a^{(1)}_{Qg})
\Big\} \ln\frac{m^2}{\mu^2} 
\nonumber \cr && \qquad
+ c^{(2)}_{2,g} + a^{(2)}_{Qg}
+ 2 \beta_0 \bar a^{(1)}_{Qg}
+ c^{(1)}_{2,q} \otimes  a^{(1)}_{Qg}
+ P^{(0)}_{qq} \otimes  \bar a^{(1)}_{Qg}
- P^{(0)}_{gg} \otimes  \bar a^{(1)}_{Qg} \,.
\nonumber \\  
\end{eqnarray}
Notice that the above formulae are still dependent on the mass factorization  
and renormalization scale $\mu^2$.
The same scale dependence was also found for the exact
expressions for the heavy quark coefficient functions 
where $Q^2$ and $m^2$ can be arbitrarily chosen.
The $\mu^2$ dependence can be attributed to coupling constant renormalization
represented by the lowest order coefficient $\beta_0$ in the
beta-function and the lowest order splitting function
$P^{(0)}_{gg}$ standing for the transition $g \rightarrow g + g$.

The computation of the asymptotic expression of the
heavy quark coefficient function corresponding to the Bethe-Heitler
process (2.13) proceeds in the same way. From (4.1)
we derive
\begin{eqnarray}
H^{(2)}_{L,q}\Big(\frac{Q^2}{m^2},\frac{\mu^2}{m^2}\Big) 
= C^{{\rm PS},(2)}_{L,q}\Big(\frac{Q^2}{\mu^2}\Big)\,,
\end{eqnarray}
\begin{eqnarray}
H^{(2)}_{2,q} \Big(\frac{Q^2}{m^2},\frac{\mu^2}{m^2}\Big) 
= A^{{\rm PS},(2)}_{Qq}\Big(\frac{\mu^2}{m^2}\Big) 
+ C^{{\rm PS},(2)}_{2,q}\Big(\frac{Q^2}{\mu^2}\Big) \,.
\end{eqnarray}
The coefficient functions $C^{{\rm PS},(2)}_{i,q}$ $(i=2,L)$ are computed in
\cite{zn} and they read
\begin{eqnarray}
C^{{\rm PS},(2)}_{L,q}\Big(\frac{Q^2}{\mu^2}\Big) = 
 \Big\{\frac{1}{2} P^{(0)}_{gq}\otimes c^{(1)}_{L,g} \Big\} 
\ln\frac{Q^2}{\mu^2} + c^{{\rm PS},(2)}_{L,q} \,,
\end{eqnarray}
\begin{eqnarray}
&&  C^{{\rm PS},(2)}_{2,q}\Big(\frac{Q^2}{\mu^2}\Big)  = 
\Big\{ \frac{1}{8} P^{(0)}_{qg}\otimes P^{(0)}_{gq} \Big\} 
\ln^2 \frac{Q^2}{\mu^2}
+ \Big\{  \frac{1}{2} P^{{\rm PS},(1)}_{qq} 
+ \frac{1}{2} P^{(0)}_{gq}\otimes c^{(1)}_{2,g} \Big\}
\ln\frac{Q^2}{\mu^2}
\nonumber  \\  && \qquad \qquad \qquad
+ c^{{\rm PS},(2)}_{2,q} \,. 
\end{eqnarray}
Using (3.31) and (4.18), (4.19) the heavy quark coefficient 
functions become
\begin{eqnarray}
&& H^{(2)}_{L,q}\Big(\frac{Q^2}{m^2},\frac{\mu^2}{m^2}\Big)   = 
\Big\{ \frac{1}{2} P^{(0)}_{gq}\otimes c^{(1)}_{L,g} \Big\}\ln\frac{Q^2}{m^2}
+ \Big\{\frac{1}{2} P^{(0)}_{gq}\otimes c^{(1)}_{L,g} \Big\}
\ln\frac{m^2}{\mu^2}
\nonumber  \\  && \qquad \qquad\qquad \qquad
+ c^{{\rm PS},(2)}_{L,q} \,, 
\end{eqnarray}
\begin{eqnarray}
&& H^{(2)}_{2,q}\Big(\frac{Q^2}{m^2}, \frac{\mu^2}{m^2}\Big) = 
\Big\{\frac{1}{8}P^{(0)}_{qg}\otimes P^{(0)}_{gq} \Big\} \ln^2\frac{Q^2}{m^2}
+\Big\{ \frac{1}{2} P^{{\rm PS},(1)}_{qq} 
+ \frac{1}{2} P^{(0)}_{gq} \otimes c^{(1)}_{2,g} \Big\} 
\nonumber \cr && \qquad \qquad \qquad
\times\ln \frac{Q^2}{m^2}
+ \Big\{ \frac{1}{4} P^{(0)}_{qg} \otimes  P^{(0)}_{gq} \Big\}
\ln\frac{Q^2}{m^2} \ln\frac{m^2}{\mu^2}
\nonumber \cr && \qquad\qquad\qquad
+\Big\{ \frac{1}{2} P^{(0)}_{gq} \otimes ( c^{(1)}_{2,g} + a^{(1)}_{Qg})
\Big\} \ln\frac{m^2}{\mu^2} 
\nonumber  \\  && \qquad \qquad \qquad
+ c^{{\rm PS},(2)}_{2,q} 
+ a^{{\rm PS},(2)}_{Qq} 
- P^{(0)}_{gq}\otimes \bar a^{(1)}_{Qg} \,.
\end{eqnarray}
Notice that the above $\mu^2$-dependence can be only attributed
to the mass factorization scheme which enters via the
transition $q \rightarrow g + q$ represented by the splitting function
$P^{(0)}_{gq}$.
The mass factorization for the heavy quark coefficient functions
where the photon couples to the light quark proceeds
in a similar way as given in eq.(4.1) i.e.,
\begin{eqnarray}
L_{i,l}\Big(\frac{Q^2}{m^2}, \frac{\mu^2}{m^2}\Big)
= A_{kl}\Big(\frac{\mu^2}{m^2}\Big)\otimes C_{i,k}
\Big(\frac{Q^2}{\mu^2}\Big)\, .
\end{eqnarray}
In the case of the Compton process (2.13) where $L_{i,q}$ is of
order $\alpha_s^2$ one can derive
\begin{eqnarray}
L^{{\rm NS},(2)}_{L,q}\Big(\frac{Q^2}{m^2}, \frac{\mu^2}{m^2}\Big)
= C^{{\rm NS},(2)}_{L,q,f}\Big(\frac{Q^2}{\mu^2}\Big) \,,
\end{eqnarray}
\begin{eqnarray}
L^{{\rm NS},(2)}_{2,q}\Big(\frac{Q^2}{m^2}, \frac{\mu^2}{m^2}\Big)
= A^{{\rm NS},(2)}_{qq,f}\Big(\frac{\mu^2}{m^2}\Big) 
+  C^{{\rm NS},(2)}_{2,q,f}\Big(\frac{Q^2}{\mu^2}\Big) \,,
\end{eqnarray}
where $C^{{\rm NS},(2)}_{i,q,f}$ $(i=2,L)$ denotes the quark loop contribution
to the non-singlet coefficient function for one specific massless
flavour $f$ which is given by (see \cite{zn})
\begin{eqnarray}
C^{{\rm NS},(2)}_{L,q,f}\Big(\frac{Q^2}{\mu^2}\Big) 
= \Big\{ - \beta_{0,f} c^{(1)}_{L,q} 
 \Big\} \ln\frac{Q^2}{\mu^2}
+ c^{{\rm NS},(2)}_{L,q,f} \,,
\end{eqnarray}
\begin{eqnarray}
&&  C^{{\rm NS},(2)}_{2,q,f}\Big(\frac{Q^2}{\mu^2}\Big)   = 
\Big\{ - \frac{1}{4} \beta_{0,f} P^{(0)}_{qq}\Big\} \ln^2 \frac{Q^2}{\mu^2}
+ \Big\{  \frac{1}{2} P^{{\rm NS},(1)}_{qq,f} - \beta_{0,f} c^{(1)}_{2,q}
\Big\} \ln\frac{Q^2}{\mu^2} 
\nonumber \\ &&  \qquad \qquad \qquad
+ c^{{\rm NS},(2)}_{2,q,f}  \,.
\end{eqnarray}
Notice that the above coefficient functions are represented
in the $\overline{\rm MS}$-scheme for the coupling constant 
renormalization (see (3.6)). However $A^{{\rm NS},(2)}_{qq}$
in (3.35) has been determined in another scheme where the heavy flavour $Q$
decouples in the strong coupling constant (see (3.7)). Choosing the
latter scheme and putting $f=Q$ we obtain instead of
(4.25), (4.26) the expressions
\begin{eqnarray}
C^{{\rm NS},(2)}_{i,q,f}\Big(\frac{Q^2}{\mu^2}, \frac{\mu^2}{m^2}\Big) = 
  C^{{\rm NS},(2)}_{i,q,f}\Big(\frac{Q^2}{\mu^2}\Big)
 + \Big\{\beta_{0,Q} 
  C^{(1)}_{i,q}\Big(\frac{Q^2}{\mu^2}\Big) \Big\}\,
 \ln \frac{m^2}{\mu^2} \,, 
\end{eqnarray}
with $C^{(1)}_{i,q}$ defined in (4.10), (4.11). Substitution of (4.27) 
and (3.35) in equations (4.23), (4.24) yields the following heavy flavour 
coefficient functions 
\begin{eqnarray}
L^{{\rm NS},(2)}_{L,q}\Big(\frac{Q^2}{m^2},\frac{\mu^2}{m^2}\Big) 
= \Big\{ - \beta_{0,Q} c^{(1)}_{L,q} \Big\}
\ln\frac{Q^2}{m^2}
+ c^{{\rm NS},(2)}_{L,q,Q} \,,
\end{eqnarray}
\begin{eqnarray}
&&  L^{{\rm NS},(2)}_{2,q}\Big(\frac{Q^2}{m^2},\frac{\mu^2}{m^2}\Big)  = 
 \Big\{ - \frac{1}{4} \beta_{0,Q} P^{(0)}_{qq}\Big\} \ln^2 \frac{Q^2}{m^2}
+\Big\{  \frac{1}{2} P^{{\rm NS},(1)}_{qq,Q} - \beta_{0,Q} c^{(1)}_{2,q}\Big\}
 \ln \frac{Q^2}{m^2}
\nonumber \\ && \qquad \qquad \qquad
+ c^{{\rm NS},(2)}_{2,q,Q}
+ a^{{\rm NS},(2)}_{qq,Q}
+ \frac{1}{4} \beta_{0,Q} \zeta(2) P^{(0)}_{qq} \,.
\end{eqnarray}
Notice that in the above expressions the $\mu^2$-dependence
has completely disappeared due to the special choice 
of the coupling constant renormalization scheme.

Since we have now all renormalization group coefficients at hand
we can calculate the heavy quark coefficient functions
$H_{k,i}$ and $L_{k,i}$ in the asymptotic limit $Q^2 \gg  m^2$ for arbitrary
$z$. The splitting functions $P^{(0)}_{ij}$, $P^{(1)}_{ij}$ and the 
nonpole terms in the OME's 
$a^{(1)}_{ij}$, $a^{(2)}_{ij}$ ($i,j = q,g$) were presented in section 3.
The coefficients
$c^{(k)}_{L,i}$, $c^{(k)}_{2,i}$ ($ k= 1,2$; $i = q,g$) 
appearing in the light quark and gluon deep inelastic coefficient functions
can be found in Appendix B of \cite{zn} (see also Appendix B in
\cite{zijl}). The analytic expressions for eqs. (4.14), (4.15),(4.20),
(4.21), (4.28), (4.29) are too long to be presented here and one can
find them in Appendix D.

Before finishing this section we want to study the behaviour of the
coefficient functions $H_{k,i}(z, Q^2/m^2, \mu^2/m^2)$
$(k = 2,L; i=q,g)$ in the limit $z\rightarrow 0$, or $\eta \rightarrow \infty$.
The behaviour of these functions for $z=0$ at arbitrary $Q^2$ has been
predicted in \cite{cch} based on the method of $k_t$-factorization.
Using the notations in \cite{rsn1} we have the following predictions
from \cite{cch}
\begin{eqnarray}
\lim_{z \rightarrow 0} 
H^{(2)}_{L,i}\Big(z, \xi,\frac{\mu^2}{m^2}\Big)
 = \frac{1}{z} 16 \pi C_i T_f \xi
\Big[ G_L(\eta,\xi) + \bar G_L(\eta,\xi) \ln\frac{\mu^2}{m^2}\Big]\,,
\end{eqnarray} 
\begin{eqnarray}
&& \lim_{z \rightarrow 0} 
H^{(2)}_{2,i}\Big(z, \xi,\frac{\mu^2}{m^2}\Big)
= \frac{1}{z} 16 \pi C_i T_f \xi
\Big[ G_T(\eta,\xi) + \bar G_L(\eta,\xi) 
\nonumber \\ && \qquad 
+\Big\{ \bar G_T(\eta,\xi) + \bar G_L(\eta,\xi) \Big\}
\ln\frac{\mu^2}{m^2}\Big]\,,
\end{eqnarray} 
with $i=q,g$ and $C_q = C_F$, $C_g = C_A$
and $\eta$, $\xi$ are defined in (2.17).
The functions $G_k(\eta,\xi)$, $\bar G_k(\eta,\xi)$ $(k=L,T)$ are
given by eqs.(19)-(24) in \cite{rsn1} for arbitrary $\xi$.
In the limit $\xi \rightarrow \infty$ they behave like
\begin{eqnarray}
\lim_{\xi \rightarrow \infty}
G_L(\eta,\xi) = \frac{1}{6\pi}
\frac{1}{\xi} [ 4 \ln \xi - \frac{4}{3}] \,,
\end{eqnarray} 
\begin{eqnarray}
\lim_{\xi \rightarrow \infty}
G_T(\eta,\xi) = \frac{1}{6\pi}
\frac{1}{\xi} [ 2\ln^2 \xi + \frac{14}{3} \ln \xi - 4 \zeta(2) + 
\frac{14}{3}] \,,
\end{eqnarray} 
\begin{eqnarray}
\lim_{\xi \rightarrow \infty}
\bar G_L(\eta,\xi) = \frac{1}{6\pi}
\frac{1}{\xi} [ - 4 ] \,,
\end{eqnarray} 
\begin{eqnarray}
\lim_{\xi \rightarrow \infty}
\bar G_T(\eta,\xi)  = \frac{1}{6\pi}
\frac{1}{\xi} [  - 4 \ln \xi + 2 ] \,.
\end{eqnarray} 
Substitution of the above equations into (4.30) and
(4.31) yields the asymptotic behaviour
\begin{eqnarray}
\lim_{z \rightarrow 0} 
H^{(2)}_{L,i}(z, \xi,\frac{\mu^2}{m^2}) = \frac{1}{z} C_i T_f 
\Big[ \frac{32}{3} \ln \xi + \frac{32}{3} \ln\frac{m^2}{\mu^2} -
\frac{32}{9} \Big]\,,
\end{eqnarray} 
\begin{eqnarray}
&&\lim_{z \rightarrow 0} 
H^{(2)}_{2,i}(z, \xi,\frac{\mu^2}{m^2}) = \frac{1}{z} C_i T_f 
\Big[ \frac{16}{3} \ln^2 \xi + \Big( \frac{32}{3} \ln \frac{m^2}{\mu^2}
+ \frac{208}{9} \Big) \ln \xi
\nonumber \\ && \qquad \qquad 
+ \frac{16}{3} \ln \frac{m^2}{\mu^2} - \frac{32}{3} \zeta(2) 
+ \frac{80}{9} \Big]\,,
\end{eqnarray} 
which agrees with expressions (D.3)-(D.6)
in the limit $z\rightarrow 0$.
\vfill
\newpage
\mysection{Results}
In this section we want to make a comparison between the exact
heavy quark coefficient functions in \cite{lrsn1} and the asymptotic
ones derived in this paper. The exact coefficient
functions which are either available in computer programs
\cite{lrsn1} or in tables \cite{rsn1} were defined in the latter
references using the notation 
$c^{(l)}_{k,i}$,  $\bar c^{(l)}_{k,i}$
$(k=2,L; i = q, \bar q, g; l=0,1)$ and
$d^{(1)}_{k,i}$  .
These twelve coefficient functions are related to the $H^{(l)}_{k,i}$
in (2.14), (2.15) and $L^{(l)}_{k,i}$ (2.16) defined in this paper as follows.
For the Born reaction we have the identity
\begin{eqnarray}
H^{(1)}_{k,g}(z,\xi)=\frac{1}{\pi} \frac{\xi}{z} c^{(0)}_{k,g}(\eta,\xi)\,,
\end{eqnarray}
with $\eta$ and $\xi$ defined in (2.17). For the gluon-fusion process
in (2.12) and the Bethe-Heitler process (2.13) we have
\begin{eqnarray}
H^{(2)}_{k,i}\Big(z,\xi, \frac{\mu^2}{m^2} \Big)=
16 \pi \frac{\xi}{z} [ 
c^{(1)}_{k,i}(\eta,\xi)\,
+ \bar c^{(1)}_{k,i}(\eta,\xi) \ln \frac{\mu^2}{m^2} ] \,,
\end{eqnarray}
and for the Compton process (2.13) one has
\begin{eqnarray}
L^{(2)}_{k,q}\Big(z,\xi, \frac{\mu^2}{m^2} \Big)=
16 \pi \frac{\xi}{z}\,  
d^{(1)}_{k,q}(\eta,\xi)\,.
\end{eqnarray}
The exact coefficient functions were plotted in figures 6-11 in 
\cite{lrsn1} as functions of $\eta$ for various values of $\xi$.
Here we want to show at which $\xi$ values the asymptotic forms of
the coefficient functions presented in section 4 give a good
approximation of the ones derived in \cite{lrsn1}.

We do not need to discuss the lowest order
coefficient functions $c^{(0)}_{L,g}$ and $c^{(0)}_{2,g}$ as they
have a very simple analytic form.  The same is true for the coefficients of the
mass factorization logarithms, $\bar c^{(1)}_{L,g}$,
$\bar c^{(1)}_{2,g}$, $\bar c^{(1)}_{L,q}$ and $\bar c^{(1)}_{2,q}$.
Therefore we concentrate on the asymptotic values of the NLO
coefficient functions and start by defining the ratios
$R^{(1)}_{k,i}$ and $T^{(1)}_{k,i}$ which are given by
\begin{eqnarray}
R^{(1)}_{k,i}(z,\xi)=\frac{c^{(1)}_{k,i}(z,\xi)}
{c^{(1),{\rm asymp}}_{k,i}(z,\xi)}\,,
\end{eqnarray}
and 
\begin{eqnarray}
T^{(1)}_{k,i}(z,\xi)=\frac{d^{(1)}_{k,i}(z,\xi)}
{d^{(1),{\rm asymp}}_{k,i}(z,\xi)}\,.
\end{eqnarray}
Here $c^{(1),{\rm asymp}}_{k,i}(z,\xi)$ 
and $d^{(1),{\rm asymp}}_{k,q}(z,\xi)$ are the asymptotic expressions 
for the heavy quark coefficient functions
derived in section 4 in the limit $\xi \rightarrow \infty$. In this paper 
the above ratios will only be presented for charm production at the HERA
collider on account of the small value of the charm mass and the $Q^2$
values accessible.
Choosing the range $ 5 < \xi < 10^{5}$  
we will study the above ratios for  
$z = 10^{-2}$ and $10^{-4}$. 

In fig.5 we show $R^{(1)}_{L,g}$ (5.4) and it
is apparent that the asymptotic formula coincides with the exact NLO
result when $\xi \ge 10^3$. There is essentially no difference between the 
ratios for $z= 10^{-2}$ and $10^{-4}$. However if $z = 10^{-4}$
the exact calculation begins to show computer instabilities when
$\xi \ge 10^4$ so we did not run at larger values.

The next figure fig.6 shows $R^{(1)}_{2,g}$ and we deduce that our
approximate formula is good for $\xi \ge 10$. This value is much lower than the
one observed for $R^{(1)}_{L,g}$ which shows that the asymptotic limit is
reached much faster for $H^{(2)}_{2,g}$ than for $H^{(2)}_{L,g}$.
Furthermore the lower bound is independent of our choice of $z$.
The scatter in fig. 6 at large values of $\xi$ reflects the numerical errors
in the computer program of the exact NLO result in \cite{rsn1}. This 
numerical uncertainty is therefore one of the reasons why we derived the
asymptotic formulae for the heavy quark coefficient functions in this paper.

Continuing with these ratios we plot $R^{(1)}_{L,q}$
versus $\xi$ in fig.7. There is not much difference between $R^{(1)}_{L,q}$
and  $R^{(1)}_{L,g}$ (see fig. 5) . The asymptotic formula for  
the quark channel is good for $\xi \ge 10^3$ for both values
of $z$. A simular observation holds for $R^{(1)}_{2,q}$ in Fig.8 when compared
with $R^{(1)}_{2,g}$ in fig. 6. Like in the latter case the exact formula
approaches the exact one  when $\xi \ge 10$.

Finally we turn to the ratios  $T^{(1)}_{k,q}$ (5.5) in Fig.9. Here we have
used the exact NLO formulae for $L^{(2)}_{k,q}$ (5.3) presented in Appendix A
rather than computing them from the program in \cite{rsn1}.
This is the reason we have no numerical troubles at very large $\xi$.
In fig. 9 and fig. 10 we have plotted $T^{(1)}_{L,q}$ and $T^{(1)}_{2,q}$
respectively. From these figures we infer that at small $z$ ( here
 $z= 10^{-4}$ ) the asymptotic formulae coincide with the exact ones over
the whole $\xi$-range. At larger $z$-values (e.g. at $z= 10^{-2}$) the
approximation gets worse and it becomes only good when $\xi \ge 10^2$.

Before drawing any conclusions about the validity of our asymptotic expressions
one has to bear the following in mind. First of all the heavy quark coefficient
functions have to be convoluted with the parton densities according to (2.8)
in order to obtain the charm contributions to the deep inelastic structure
functions. Therefore there are parts of the integration region over $z$ which
can become more important than some other ones. Second the lower bounds on
the $\xi$-values given above have to be viewed in the mathematical rather
than in the physical sense. Experimentally the structure function
$F_2(x,Q^2,m_c^2)$ will become much better determined than the longitudinal
one given by $F_L(x,Q^2,m_c^2)$. In the latter case we are already glad that
it can be measured up to $20-30\%$ accuracy. Furthermore in this paper we
are dealing with NLO corrections which have to be less precise than the
Born approximation. Also the contribution coming from the Compton subprocess
leading to the plots in figs. 9,10 is much smaller than the one originating
from the photon-gluon fusion reaction for which the ratios $R^{(1)}_{k,g}$
are plotted in figs. 5,6. Therefore the physical bounds on $\xi$ can be put
much smaller than the ones given above. In view of the theoretical and
experimental uncertainties we can state that the asymptotic formulae for
the heavy flavour coefficient functions can be used when $\xi \ge 4$ in the
case of $F_2$ whereas $\xi \ge 10$ is good enough for $F_L$.

To summarize the calculations presented in this paper we have used the OPE
techniques and the renormalization group equations to find analytic
formulae for the asymptotic behaviour ($\xi \gg 1$) of the heavy flavour 
coefficient functions which enter in deep inelastic electroproduction.
We have tested these asymptotic formulae against the exact NLO
results in our rather complicated computer programs and find good agreement
when $\xi \ge 4$ for $F_2$ and $\xi \ge 10$ in the case of $F_L$.
Below these values our asymptotic formulae fail and one has to use
the exact NLO computer programs to compute the heavy quark
coefficient functions.

Acknowledgements.

The work of Y. Matiounine and J. Smith was supported in 
part under the contract NSF 93-09888.
R. Migneron would like to thank the Netherlands Organization for Scientific
Research (NWO) for financial support.
\vfill
\newpage
\mysection*{Appendix A}
\setcounter{section}{1}
In this Appendix we present the exact expressions for the heavy quark
coefficient functions $L^{(2)}_{i,q}$ (2.16) corresponding to the
Compton reaction (2.13). The calculation is straightforward because 
one can first integrate over the heavy quark momenta in the final state 
without affecting the momentum of the remaining light quark 
(see figs. 5c,d in \cite{lrsn1}).
The phase space integrals are then the same as the ones obtained for the
process $\gamma^*+q\rightarrow g^*+q $ $ (g^*\rightarrow Q+\overline{Q})$
where the gluon $g^*$ becomes virtual. After integration over the virtual 
mass of the gluon one gets the expressions 
\begin{eqnarray}
&&L^{(2)}_{L,q}(z,Q^2,m^2)=C_FT_f\Big[96z(\frac{z}{\xi})^2\Big\{
\ln\Big(\frac{1-z}{z^2}\Big)L_1+L_1L_2+2(-DIL_1
\nonumber \\ && \qquad
+DIL_2+DIL_3-DIL_4)\Big\}
+\Big(\frac{z}{(1-z)\xi }\Big)^2(64-288z+192z^2)L_1
\nonumber \\ && \qquad
+z\Big\{\frac{16}{3}-\frac{416}{3}\Big(\frac{z}{\xi }\Big)+\frac{1408}{3}
\Big(\frac{z}{\xi }\Big)^2\Big\}\frac{L_{3}}{sq_2}+
\Big\{\frac{16}{3}-\frac{400}{18}z+\Big(\frac{z}{(1-z)\xi }\Big)
\nonumber \\ && \qquad
\times (-\frac{160}{3}
+\frac{3824}{9}z-\frac{992}{3}z^2)\Big\}sq_1\Big] \,,
\end{eqnarray}
\begin{eqnarray}
 && L^{(2)}_{2,q}(z,Q^2,m^2)=C_FT_f\Big[\Big\{\frac{4}{3}\frac{1+z^2}{1-z}
-\frac{16}{1-z}\Big(\frac{z}{\xi }\Big)^2(1-9z+9z^2)\Big\}
\nonumber \\ && \qquad
\times \Big\{
\ln\Big(\frac{1-z}{z^2}\Big)L_1
+L_1L_2+2(-DIL_1
+DIL_2+DIL_3-DIL_4)\Big\}
\nonumber \\ && \qquad
+\Big\{-\frac{8}{3}+\frac{4}{1-z}+\Big(\frac{z}{(1-z)\xi }\Big)^2
\Big(128-432z+288z^2-\frac{8}{1-z}\Big)\Big\}L_1
\nonumber \\ && \qquad
+\Big\{\frac{88}{9}+\frac{136}{9}z-\frac{152}{9}\frac{1}{1-z}
+\Big(\frac{z}{(1-z)\xi }\Big)\Big(\frac{464}{9}-\frac{512}{3}z
+ \frac{2048}{9}z^2\Big)
\nonumber \\ && \qquad
+\Big(\frac{z}{(1-z)\xi }\Big)^2\Big(-\frac{832}{9}+\frac{6208}{9}z
-\frac{11392}{9}z^2+\frac{6016}{9}z^3\Big)\Big\}\frac{L_{3}}{sq_2}
\nonumber \\ && \qquad
+\Big\{-\frac{272}{27}
-\frac{1244}{27}z+\frac{718}{27}\frac{1}{1-z}+
\Big(\frac{z}{(1-z)\xi }\Big)\Big(-\frac{3424}{27}+\frac{15608}{27}z
\nonumber \\ && \qquad
-\frac{4304}{9}z^2
+\frac{20}{27}\frac{1}{1-z}\Big)\Big\}sq_1\Big] \,,
\end{eqnarray}
where $\xi =Q^2/m^2$ (2.17). Further we have defined
\begin{eqnarray}
sq_1=\sqrt{1-4\frac{z}{\xi}} 
\qquad , \qquad sq_2=\sqrt{1-4\frac{z}{(1-z)\xi}} \, ,
\end{eqnarray}
\begin{eqnarray}
L_1=\ln\Big(\frac{1+sq_1}{1-sq_1}\Big) 
\quad , \quad L_2=\ln\Big(\frac{1+sq_2}{1-sq_2}\Big)  
\quad , \quad L_3=\ln\Big(\frac{sq_2+sq_1}{sq_2-sq_1}\Big) \,,
\end{eqnarray}
\begin{eqnarray}
DIL_1={\rm Li}_2\Bigg(\frac{(1-z)(1+sq_1)}{1+sq_2}\Bigg) 
\qquad , \qquad DIL_2={\rm Li}_2\Big(\frac{1-sq_2}{1+sq_1}\Big) \,,
\end{eqnarray}
\begin{eqnarray}
DIL_3={\rm Li}_2\Big(\frac{1-sq_1}{1+sq_2}\Big) 
 \qquad , \qquad DIL_4={\rm Li}_2\Big(\frac{1+sq_1}{1+sq_2}\Big) \, ,
\end{eqnarray}
with $0 < z < \xi/(\xi+4)$.
\vfill
\newpage
\mysection*{Appendix B}
\setcounter{section}{2}
Here we apply the method of integration by parts which enables
us to reduce scalar two-loop OME integrals with 
five different propagators to expressions containing only four different 
propagators. The method is a generalization of the trick invented in
\cite{ct} where it was used to reduce two-loop self-energy integrals
with five different propagators to integrals which only contain four
different propagators.

Feynman integrals with five different propagators emerge
from the computation of the graphs $e,f,h,l,m,n,o$ in fig. 2.
Let us first start with graphs 2l, 2o which lead to the following
expression
\begin{eqnarray}
I^{(m)} = \int \, \frac{d^n q}{(2\pi)^n} \int \, \frac{d^n k}{(2\pi)^n} 
\frac{(\Delta\cdot q)^m}{D_k D_q^a D_{kp} D_{qp} D_{kq}} \,,
\end{eqnarray}
where $a$ is 1 or 2. The denominators are defined as
\begin{eqnarray}
D_k = k^2 - m_1^2 \,,
\end{eqnarray}
\begin{eqnarray}
D_q = q^2 - m_2^2 \,,
\end{eqnarray}
\begin{eqnarray}
D_{kp} = (k-p)^2 - m_3^2 \,,
\end{eqnarray}
\begin{eqnarray}
D_{qp} = (q-p)^2 - m_4^2\,,
\end{eqnarray}
\begin{eqnarray}
D_{kq} = (k-q)^2 - m_5^2 \,.
\end{eqnarray}
In $n$-dimensional regularization the following integral is zero
\begin{eqnarray}
0 = \int \, \frac{d^n q}{(2\pi)^n} \int \, \frac{d^n k}{(2\pi)^n} 
\frac{\partial}{\partial k_\mu} \Big[\frac{k_\mu 
(\Delta\cdot q)^m}{D_k D_q^a D_{kp} D_{qp} D_{kq}} \Big] \,.
\end{eqnarray}
Differentiating the right-hand-side of (B.7) we obtain
\begin{eqnarray}
&& 0 =  n I^{(m)} + \int\, \frac{d^n q}{(2\pi)^n}\int\,\frac{d^n k}{(2\pi)^n} 
\frac{(\Delta\cdot q)^m}{D_k D_q^a D_{kp} D_{qp} D_{kq}}
\nonumber \\ &&\qquad
\Big\{ - \frac{2k^2}{D_k} - \frac{2k\cdot(k-p)}{D_{kp}}
- \frac{2k\cdot (k-q)}{D_{kq}} \Big\} \,.
\end{eqnarray}
In the case that the following conditions are satisfied i.e.,
$ m_1 = 0$, $p^2 = m_1^2 + m_3^2$, $m_2^2 =  m_1^2 + m_5^2$ one can write
\begin{eqnarray}
-2k\cdot (k-p) = - D_k - D_{kp} \,,
\end{eqnarray}
\begin{eqnarray}
-2k\cdot (k-q) = - D_k - D_{kq} + D_q \,.
\end{eqnarray}
Substitution of (B.9) and (B.10) into (B.8) yields
\begin{eqnarray}
&&I^{(m)}=\frac{1}{n-4}\int\,\frac{d^n q}{(2\pi)^n}\int\,\frac{d^n k}{(2\pi)^n} 
(\Delta\cdot q)^m
\Big\{ \frac{1}{ D_q^a D_{kp}^2 D_{qp} D_{kq}}
\nonumber \\ && \qquad 
+  \frac{1}{D_q^a D_{kp} D_{qp} D_{kq}^2}
- \frac{1}{D_k D_q^{a-1} D_{kp} D_{qp} D_{kq}^2} \Big\}\,.
\end{eqnarray}
If $a=1$ we have already reached our goal. When $a=2$ we have to repeat
the trick for the last term in (B.11) so that we finally get
\begin{eqnarray}
&&I^{(m)}=\frac{1}{n-4}\int\,\frac{d^n q}{(2\pi)^n}\int\,\frac{d^n k}{(2\pi)^n} 
(\Delta\cdot q)^m
\Big[ \frac{1}{ D_q^a D_{kp}^2 D_{qp} D_{kq}}
\nonumber \\ && \qquad
+  \frac{1}{D_q^a D_{kp} D_{qp} D_{kq}^2}
- \frac{1}{n-5}
\Big\{\frac{1}{D_q^{a-1} D_{kp}^2 D_{qp} D_{kq}^2} 
\nonumber \\ && \qquad
+ \frac{2}{D_q^{a-1} D_{kp} D_{qp} D_{kq}^3} 
-\frac{2}{D_k D_q^{a-2} D_{kp} D_{qp} D_{kq}^3} 
\Big\} \Big]\,.
\end{eqnarray}
For graphs $2e,2h$ we have to differentiate as follows
\begin{eqnarray}
&& 0 = \int \, \frac{d^n q}{(2\pi)^n} \int \, \frac{d^n k}{(2\pi)^n} 
\frac{\partial}{\partial k_\mu} \Big[\frac{(k-q)_\mu 
(\Delta\cdot q)^m}{D_k D_q^a D_{kp} D_{qp} D_{kq}} \Big]
\nonumber \\ &&\qquad
=  n I^{(m)} + \int \, \frac{d^n q}{(2\pi)^n} \frac{d^n k}{(2\pi)^n} 
\frac{(\Delta\cdot q)^m}{D_k D_q^a D_{kp} D_{qp} D_{kq}} \,.
\nonumber \\ &&\qquad
\times \Big\{ - \frac{2k\cdot(k-q)}{D_k} - \frac{2(k-q)\cdot(k-p)}{D_{kp}}
- \frac{2(k-q)^2}{D_{kq}} \Big\} \,.
\end{eqnarray}
Imposing the following conditions $m_5=0$, $m_2^2 = m_1^2 + m_5^2$,
$m_4^2 = m_3^2 +m_5^2$ one gets identity (B.10) and
\begin{eqnarray}
-2(k-q) \cdot (k-p) = D_{qp} - D_{kq} - D_{kp}
\end{eqnarray}
so that $I^{(m)}$ equals
\begin{eqnarray}
&&I^{(m)}=\frac{1}{n-4}\int\,\frac{d^n q}{(2\pi)^n}\int\,\frac{d^n k}{(2\pi)^n} 
(\Delta\cdot q)^m
\Big[ \frac{1}{ D_k^2 D_q^a D_{kp} D_{qp} }
\nonumber \\ &&
+  \frac{1}{D_k D_q^a D_{kp}^2 D_{qp} }
-  \frac{1}{D_k D_q^a D_{kp}^2 D_{kq} } 
-\frac{1}{D_k^2 D_q^{a-1} D_{kp} D_{qp} D_{kq}} 
\Big]\,.
\end{eqnarray}
For $a=1$ we have already four different propagators. If $a=2$ one
has to repeat the trick again analogous to (B.11).

The trick of integration by parts also applies to graph $2m$.
For this graph we have the integral
\begin{eqnarray}
J^{(m)}=\int \, \frac{d^n q}{(2\pi)^n}\int \, \frac{d^n k}{(2\pi)^n} 
\frac{(\Delta\cdot q)^m}{D_k D_q^a D_{kp} D_{kq} D_{kqp}} \,,
\end{eqnarray}
with
\begin{eqnarray}
D_{kqp} = (k-q-p)^2 - m_6^2\,.
\end{eqnarray}
Performing the differentiation with respect to $k$ in the same way
as in (B.7) one obtains
\begin{eqnarray}
&&J^{(m)}=\frac{1}{n-4}\int\,\frac{d^n q}{(2\pi)^n}\int\,\frac{d^n k}{(2\pi)^n} 
(\Delta\cdot q)^m
\Big[ \frac{1}{ D_q^a  D_{kp}^2 D_{kq} D_{kpq} }
+  \frac{1}{ D_q^a D_{kp} D_{kq}^2 D_{kqp} }
\nonumber \\ && \qquad
+  \frac{1}{ D_k D_q^a D_{kp} D_{kpq}^2 } 
+\frac{1}{D_k D_q^a D_{kq} D_{kpq}^2 } 
\nonumber \\ && \qquad
-\frac{1}{D_k D_q^{a-1} D_{kp} D_{kq}^2 D_{kqp}} 
- \frac{1}{D_k D_q^{a-1} D_{kq} D_{kp} D_{kpq}^2 } 
\Big]\,.
\end{eqnarray}
In the above expression we have used the relations in (B.9), (B.10)
and
\begin{eqnarray}
-2k\cdot (k-q-p) = D_q - D_{kq} - D_{kp} \,,
\end{eqnarray}
which only holds under the conditions $p^2 = m_3^2$ , $m_1^2 = 0$,
$m_2^2 = m_5^2$.

The integrals corresponding to graphs $2f$ and $2n$ need some special treatment
because of the sum
$\sum^m_{i=0} (\Delta\cdot k)^i (\Delta \cdot q)^{m-i}$ which appears
in the operator vertex. Notice that this sum also appears in graphs $2h, 2o$
but after some rearrangement of terms it drops out of the integrals. 
For graph $2n$  we have the expression
\begin{eqnarray}
K^{(m)}=\int \, \frac{d^n q}{(2\pi)^n}\int\, \frac{d^n k}{(2\pi)^n} 
\frac{(\Delta\cdot k)^m - (\Delta\cdot q)^m}{(\Delta\cdot k - \Delta \cdot q)}
\frac{1}{D_q D_k D_{kp} D_{kq} D_{kqp}} \,,
\end{eqnarray}
with $m_1^2 = m_2^2 = m_3^2 = m^2$ and $m_5^2 = m_6^2 = 0$.
The integral $K^{(m)}$ can be split into
$K^{(m)} = K^{(m)}_1 - K^{(m)}_2$ with
\begin{eqnarray}
K^{(m)}_1 = \int \, \frac{d^n q}{(2\pi)^n}\int\, \frac{d^n k}{(2\pi)^n} 
\frac{(\Delta\cdot k)^m }{(\Delta\cdot k - \Delta \cdot q)}
\frac{1}{D_q D_k D_{kp} D_{kq} D_{kqp}}\,,
\end{eqnarray}
\begin{eqnarray}
K^{(m)}_2 = \int \, \frac{d^n q}{(2\pi)^n}\int\, \frac{d^n k}{(2\pi)^n} 
\frac{(\Delta\cdot q)^m}{(\Delta\cdot k - \Delta \cdot q)}
\frac{1}{D_q D_k D_{kp} D_{kq} D_{kqp}}\,.
\end{eqnarray}
Differentiating the integrand in (B.21) with respect to $q$
and (B.22) with respect to $k$ analogous to (B.13) we obtain
\begin{eqnarray}
&&K^{(m)}_1=\frac{1}{n-5}\int\,\frac{d^n q}{(2\pi)^n}
\int\,\frac{d^n k}{(2\pi)^n} 
\frac{(\Delta\cdot k)^m}{(\Delta\cdot k - \Delta\cdot q)}
\Big[ \frac{1}{ D_k D_q^2 D_{kp} D_{kqp} }
\nonumber \\ &&\qquad
+  \frac{1}{D_k D_q D_{kp}D_{kqp}^2 }
-  \frac{1}{D_q^2 D_{kp} D_{kq} D_{kpq} } 
\Big]\,,
\end{eqnarray}
\begin{eqnarray}
&&K^{(m)}_2 =\frac{1}{n-5}\int \,\frac{d^n q}{(2\pi)^n}
\int\,  \frac{d^n k}{(2\pi)^n} 
\frac{(\Delta\cdot q)^m}{(\Delta\cdot k - \Delta \cdot q)}
\Big[ \frac{1}{ D_k^2 D_q D_{kp} D_{kqp} }
\nonumber \\ &&\qquad
+  \frac{1}{D_q D_{kp}^2 D_{kq} D_{kqp}}
+  \frac{1}{D_k D_q D_{kp}^2 D_{kq} } 
+ \frac{1}{D_k D_q D_{kp} D_{kqp}^2} 
\nonumber \\ &&\qquad
-\frac{1}{D_k^2 D_{kp} D_{kq} D_{kqp}} 
-\frac{1}{D_k D_{kp}^2 D_{kq} D_{kqp}} 
\Big]\,.
\end{eqnarray}
As $p^2 =0$ the last two terms in (B.24) are equal to zero. 

Finally we apply the method of partial integration to graph $2f$.
The corresponding integral becomes
\begin{eqnarray}
L^{(m)} = \int \, \frac{d^n q}{(2\pi)^n}\int\, \frac{d^n k}{(2\pi)^n} 
\frac{(\Delta\cdot k)^m - (\Delta\cdot q)^m}{(\Delta\cdot k - \Delta \cdot q)}
\frac{1}{D_k D_q D_{kp} D_{qp} D_{kq}}\,,
\end{eqnarray}
with $m_1^2 = m_2^2 = m_3^2 = m_4^2=m^2$ and $m_5^2=0$.
$L^{(m)}$ can be split in the same way as done for $K^{(m)}$ in (B.20).
However because of the symmetry in $k\leftrightarrow q$ one can simplify
the calculation. Here we have
\begin{eqnarray}
L^{(m)} = 2 L^{(m)}_1  = 2 \int \, \frac{d^n q}{(2\pi)^n} 
\int \, \frac{d^n k}{(2\pi)^n} 
\frac{(\Delta\cdot k)^m }{(\Delta\cdot k - \Delta \cdot q)}
\frac{1}{D_k D_q D_{kp} D_{qp} D_{kq}} \!.
\end{eqnarray}
Differentiating the integrand with respect to $q$ analogous to (B.13)
we obtain
\begin{eqnarray}
&&L^{(m)}_1 =\frac{1}{n-5}\int \,\frac{d^n q}{(2\pi)^n} 
\int \, \frac{d^n k}{(2\pi)^n} 
\frac{(\Delta\cdot k)^m }{(\Delta\cdot k - \Delta \cdot q)}
\Big[ \frac{1}{ D_k D_q^2 D_{kp} D_{qp} }
\nonumber \\ && \qquad
+  \frac{1}{D_k D_q D_{kp} D_{qp}^2 }
- \frac{1}{D_q^2 D_{kp} D_{qp} D_{kq}} 
- \frac{1}{D_k D_q  D_{qp}^2 D_{kq}} 
 \Big]\,.
\end{eqnarray}
\vfill
\newpage
\mysection*{Appendix C}
\setcounter{section}{3}
Here we present the unrenormalized operator matrix elements 
$\hat A_{ij}^{(2)}$ whose general structure expressed in 
renormalization group coefficients was derived in section 3. 
After having carried out mass
renormalization the two-loop OME in fig. 2 is given by the
following expression ( see also (3.22) )
\begin{eqnarray}
&&\hat A_{Qg}^{(2)}\Big(\frac{\mu^2}{m^2},\epsilon\Big)
=S_{\epsilon}^2\Big(\frac{m^2}
{\mu^2}\Big)^{\epsilon}
\Big[\frac{1}{\epsilon^2}\Big\{C_FT_f[ (32 -64 z+64 z^2)\ln(1-z)
\nonumber \\ && \qquad
-(16 -32 z+ 64 z^2)\ln z -(8 - 32 z)]
\nonumber \\ && \qquad
+C_AT_f\Big[-(32 - 64 z + 64 z^2)\ln(1-z)
-(32 + 128 z)\ln z-\frac{64}{3z}
\nonumber \\ && \qquad
-16  - 128z+\frac{496}{3}z^2\Big]\Big\}
\nonumber \\ && \qquad
+\frac{1}{\epsilon}\Big\{C_FT_f \Big[( 8 - 16 z + 16z^2)[2\ln z\ln(1-z)
-\ln^2(1-z)+2\zeta(2)]
\nonumber \\ && \qquad
-(4 - 8 z +16 z^2)\ln^2z-32z(1-z)\ln(1-z)
\nonumber \\ && \qquad
-(12 - 16 z + 32 z^2)\ln z  - 56+116z -80z^2 \Big]
\nonumber \\ && \qquad
+ C_AT_f\Big[(16 +32 z +32 z^2)[{\rm Li}_2(-z) + \ln z\ln(1+z) ]
\nonumber \\ && \qquad
+(8 - 16 z + 16 z^2)\ln^2(1-z)
+(8 + 16 z)\ln^2z
\nonumber \\ && \qquad
+32z\zeta(2)+32z(1-z)\ln(1-z)
-\Big(8+64z+\frac{352}{3}z^2\Big)\ln z
\nonumber \\ && \qquad
-\frac{160}{9z}+16-200z+\frac{1744}{9}z^2 \Big]\Big\}
+a^{(2)}_{Qg}(z)\Big]
\nonumber \\ && \qquad
+\sum_{H=c,b,t}S_{\epsilon}^2\Big(
\frac{m_H^2}{\mu^2}\Big)^{\epsilon/2} \Big( \frac{m^2}{\mu^2}\Big)^{\epsilon/2}
\Big[\frac{1}{\epsilon^2}T_f^2\Big(
-\frac{64}{3}+\frac{128}{3}z-\frac{128}{3}z^2\Big)
\nonumber \\ && \qquad \qquad
\times \Big(1 + \frac{\epsilon^2}{4}\zeta(2)\Big)\Big] \,,
\end{eqnarray}
with 
\begin{eqnarray}
&&a^{(2)}_{Qg}(z)=C_FT_f\Big\{(1-2z+2z^2)[8\zeta(3)+8\zeta(2)\ln(1-z)
+\frac{4}{3}\ln^3(1-z)
\nonumber \\ && \qquad
-8\ln(1-z){\rm Li}_2(1-z)
+4\zeta(2)\ln z
-4\ln z\ln^2(1-z)
\nonumber \\ && \qquad
+\frac{2}{3}\ln^3z
-8\ln z{\rm Li}_2(1-z)
+8{\rm Li}_3(1-z)
-24{\rm S}_{1,2}(1-z)]
\nonumber \\ && \qquad
+z^2\Big[-24\zeta(2)\ln z+\frac{4}{3}\ln^3z
+16\ln z{\rm Li}_2(1-z)+32{\rm S}_{1,2}(1-z)\Big]
\nonumber \\ && \qquad
-(4+96z-64z^2){\rm Li}_2(1-z)
-(6-56z+40z^2)\zeta(2)
\nonumber \\ && \qquad
-(8+48z-24z^2)\ln z\ln(1-z)
+(4+8z-12z^2)\ln^2(1-z)
\nonumber \\ && \qquad
-(1+12z-20z^2)\ln^2z-(52z-48z^2)\ln(1-z)
\nonumber \\ && \qquad
-(16+18z+48z^2)\ln z
+26-82z+80z^2\Big\}
\nonumber \\ && \qquad
+C_AT_f\Big\{(1-2z+2z^2) [-8\zeta(2)\ln(1-z)
-\frac{4}{3} \ln^3(1-z)
\nonumber \\ && \qquad
+8\ln(1-z){\rm Li}_2(1-z)-8{\rm Li}_3(1-z)]
+(1+2z+2z^2)
\nonumber \\ && \qquad
\times [-8\zeta(2)\ln(1+z)
-16\ln(1+z){\rm Li}_2(-z)
-8\ln z\ln^2(1+z)
\nonumber \\ && \qquad
+4\ln^2z\ln(1+z) + 8\ln z{\rm Li}_2(-z)-8{\rm Li}_3(-z)
-16{\rm S}_{1,2}(-z)]
\nonumber \\ && \qquad
+(16+64z)[2{\rm S}_{1,2}(1-z)
+\ln z{\rm Li}_2(1-z)]
-\Big(\frac{4}{3} +  \frac{8}{3} z\Big)\ln^3z
\nonumber \\ && \qquad
+(8-32z+16z^2)\zeta(3)-(24+96z)\zeta(2)\ln z+(16+16z^2)
\nonumber \\ && \qquad
\times [ {\rm Li}_2(-z) + \ln z\ln(1+z)  ]
+\Big(\frac{32}{3z}+12+64z-\frac{272}{3}z^2\Big)
{\rm Li}_2(1-z)
\nonumber \\ && \qquad
-\Big( 16 + 80 z - 128 z^2+\frac{16}{z}\Big)\zeta(2)
-4z^2\ln z\ln(1-z)
\nonumber \\ && \qquad
-(2+8z-10z^2)\ln^2(1-z)+\Big(2+8z+\frac{46}{3}z^2\Big)\ln^2z
\nonumber \\ && \qquad
+(4+16z-16z^2)\ln(1-z)
-\Big(\frac{56}{3}+\frac{172}{3}z+\frac{1600}{9}z^2\Big)\ln z
\nonumber \\ && \qquad
-\frac{448}{27z}-\frac{4}{3}-\frac{628}{3}z
+\frac{6352}{27}z^2\Big\} \,.
\end{eqnarray}
The unrenormalized OME corresponding to fig.3 ( see (3.29) )
is given by
\begin{eqnarray}
&&\hat A^{{\rm PS},(2)}_{Qq}\Big(\frac{\mu^2}{m^2},\epsilon\Big)=S_{\epsilon}^2
\Big(\frac{m^2}{\mu^2}\Big)^{\epsilon}C_FT_f\Big\{
\frac{1}{\epsilon^2}\Big(-32(1+z)\ln z-\frac{64}{3z}-16
\nonumber \\ && \qquad
+ 16 z +\frac{64}{3}z^2\Big)
+\frac{1}{\epsilon}\Big(8(1+z)\ln^2z-\Big(8+40z+\frac{64}{3}z^2\Big)\ln z
-\frac{160}{9z}
\nonumber \\ && \qquad \qquad 
+16-48z+\frac{448}{9}z^2\Big)
+a^{{\rm PS},(2)}_{Qq}(z)\Big\} \,,
\end{eqnarray}
with
\begin{eqnarray}
&&a^{{\rm PS},(2)}_{Qq}(z)=
\Big[ (1+z)\Big(32{\rm S}_{1,2}(1-z)+16\ln z{\rm Li}_2(1-z)
-24\zeta(2)\ln z
\nonumber \\ && \qquad
-\frac{4}{3}\ln^3z\Big)
+\Big(\frac{32}{3z}+8-8z-\frac{32}{3}z^2\Big) {\rm Li}_2(1-z)
\nonumber \\ && \qquad
+ \Big( -\frac{16}{z}-12+12z+16z^2\Big)\zeta(2)
+\Big(2+10z+\frac{16}{3}z^2\Big)  \ln^2z
\nonumber \\ && \qquad
-\Big(\frac{56}{3}+\frac{88}{3}z
+\frac{448}{9}z^2\Big)\ln z
-\frac{448}{27z} - \frac{4}{3}
-\frac{124}{3}z+\frac{1600}{27}z^2 \Big]  \,.
\end{eqnarray}
The unrenormalized OME corresponding to fig.4 is (see (3.33)) 
\begin{eqnarray}
&& \hat A^{{\rm NS},(2)}_{qq,Q}\Big(\frac{\mu^2}{m^2},\epsilon\Big)
=S_{\epsilon}^2\Big(
\frac{m^2}{\mu^2}\Big)^{\epsilon} C_F T_f \Big\{\frac{1}{\epsilon^2}\Big[
\frac{32}{3}\Big(\frac{1}{1-z}\Big)_+
-\frac{16}{3}-\frac{16}{3}z+8\delta(1-z)\Big]
\nonumber \\ && \qquad
+\frac{1}{\epsilon}\Big[\frac{80}{9}\Big(\frac{1}{1-z}\Big)_+ +\frac{8}{3}
\frac{1+z^2}{1-z}\ln z+\frac{8}{9}-\frac{88}{9}z
+\delta(1-z)\Big(
\frac{16}{3}\zeta(2)+\frac{2}{3}\Big)\Big]
\nonumber \\ && \qquad
+a^{{\rm NS},(2)}_{qq,Q}(z)\Big\} \,,
\end{eqnarray}
with
\begin{eqnarray}
 && a^{{\rm NS},(2)}_{qq,Q}(z)= \Big(\frac{8}{3}\Big(\frac{1}{1-z}\Big)_+ 
-\frac{4}{3} - \frac{4}{3} z\Big) \zeta(2)
%
%
+\frac{1+z^2}{1-z}\Big(\frac{2}{3}\ln^2z+\frac{20}{9}\ln z\Big)
\nonumber \\ && \qquad
+\frac{8}{3}(1-z)\ln z 
+\frac{224}{27}\Big(\frac{1}{1-z}\Big)_+
+\frac{44}{27}-\frac{268}{27}z
\nonumber \\ && \qquad
+\delta(1-z)\Big(-\frac{8}{3}\zeta(3)+\frac{58}{9}
\zeta(2)+\frac{73}{18}\Big) \,.
\end{eqnarray}
Here the term $1/(1-z)_+$ has to be defined in the distribution sense
\begin{eqnarray}
\int^1_0 \,dz \Big(\frac{1}{1-z}\Big)_+ f(z)
=\int_0^1dz\frac{1}{1-z}[f(z)-f(1)] \,.
\end{eqnarray}
Notice that as long as $z<Q^2/(Q^2+4m^2)$ with $m^2\neq 0$ the terms
proportional to $\delta(1-z)$ do not contribute and the subscript $+$
in $(1/(1-z))_+$ can be dropped.
We will comment on this in Appendix D after having predicted the asymptotic
expression for $L^{{\rm NS},(2)}_{2,q}(z,Q^2,m^2)$. The OME's which emerge from
the calculation of graphs $1-4$ usually appear in the Mellin
transformed representation
\begin{eqnarray}
A^{(m)}_{ij}=\frac{1}{2}[1+(-1)^m]\int_0^1 dz~z^{m-1}A_{ij}(z) \, .
\end{eqnarray}
This implies that the anomalous dimensions
$\gamma_{ij}^{(m)}$ correspond to the physical operators listed in
eqs.(2.24)-(2.26)
only for even $m$. 
\vfill
\newpage
\mysection*{Appendix D}
\setcounter{section}{4}
In this appendix we present the heavy quark coefficient functions
$H^{(2)}_{i,j}$ and $L^{(2)}_{i,j}$ $(i=L,2;j=q,g)$ in the asymptotic
limit $Q^2 \gg m^2$. Starting with process (2.11) the heavy quark coefficient
functions read ( see (4.6), (4.7) )
\begin{eqnarray}
 H^{(1)}_{L,g}(z,Q^2,m^2)=T_f [ 16 z (1-z) ] \,,
\end{eqnarray}
\begin{eqnarray}
 && H^{(1)}_{2,g}(z,Q^2,m^2)=T_f[(4-8z+8z^2)\Big(\ln\frac{Q^2}{m^2} 
+ \ln(1-z) - \ln z\Big)
\nonumber \\ && \qquad
-4+32z-32z^2] \,.
\end{eqnarray}
In next-to-leading order the coefficient functions given by process 
(2.12) and expressions (4.14), (4.15) read
\begin{eqnarray}
 && H^{(2)}_{L,g}(z,Q^2,m^2)=
\Big[ C_F T_f \{ 32 z \ln z + 16 + 16z - 32 z^2 \}
\nonumber \\ && \qquad
+ C_A T_f \Big\{ 64 z (1-z) \ln(1-z)
- 128 z \ln z + \frac{32}{3z}  - 32 
\nonumber \\ && \qquad
- 160 z + \frac{544}{3} z^2\Big\}\Big]
\ln \frac{Q^2}{m^2} 
\nonumber \\ && \qquad
+ C_A T_f \Big\{ 64 z (1-z) \ln (1-z) - 128 z \ln z + \frac{32}{3z} - 32 
\nonumber \\ && \qquad
 - 160 z + \frac{544}{3} z^2 \Big\} \ln \frac{m^2}{\mu^2}
\nonumber \\ && \qquad
+ C_F T_f \Big\{ \Big( \frac{64}{15z^2}
- \frac{64}{3} z + \frac{128}{5} z^3\Big) [ {\rm Li}_2(-z) + \ln z \ln (1+z)]
\nonumber \\ && \qquad
+32 z [ {\rm Li}_2(1-z) + \ln z \ln (1-z)]
-\Big(\frac{64}{3} z - \frac{128}{5}z^3\Big) \zeta(2) 
\nonumber \\ && \qquad
-\Big(\frac{64}{3} z + \frac{64}{5} z^3\Big) \ln^2 z
+( 16 + 48 z - 64 z^2) \ln (1-z)
\nonumber \\ && \qquad
+ \Big(- \frac{64}{15z} - \frac{208}{15} 
- \frac{416}{5}z + \frac{192}{5}z^2\Big) \ln z
+ \frac{64}{15z} - \frac{256}{15} - \frac{608}{5}z + \frac{672}{5} z^2\Big\}
\nonumber \\ && \qquad
+ C_A T_f \Big\{ ( 64 z + 64 z^2)[{\rm Li}_2(-z) + \ln z \ln (1+z)]
- 128 z {\rm Li}_2(1-z) 
\nonumber \\ && \qquad
+ 64 z^2 \zeta(2) 
- ( 192 z - 64 z^2 ) \ln z \ln (1-z) + (32 z - 32 z^2) \ln^2 (1-z)
\nonumber \\ && \qquad
+ 96 z \ln ^2 z
+\Big( \frac{32}{3z} - 32 - 288 z + \frac{928}{3} z^2\Big) \ln (1-z)
\nonumber \\ && \qquad
+(32 + 256 z - 416 z^2) \ln z
- \frac{32}{9z} + \frac{32}{3} + \frac{544}{3} z - \frac{1696}{9}z^2\Big\} \,,
\end{eqnarray}
\begin{eqnarray}
 && H^{(2)}_{2,g}(z,Q^2,m^2)=
\Big[C_F T_f \{ ( 8 - 16 z +16 z^2) \ln (1-z) 
\nonumber \\ && \qquad
- ( 4 - 8z +16 z^2) \ln z  - 2 + 8z\}
\nonumber \\ && \qquad
+ C_A T_f \{ ( 8 - 16 z + 16 z^2) \ln (1-z)
+ (8+32 z) \ln z + \frac{16}{3z} 
\nonumber \\ && \qquad
+ 4 + 32 z - \frac{124}{3} z^2 \}\Big] 
\ln^2 \frac{Q^2}{m^2}
\nonumber \\ && \qquad
+\Big[ C_F T_f \{ (8-16 z) {\rm Li}_2(1-z)
- ( 32 - 64 z + 64 z^2) \zeta(2) 
\nonumber \\ && \qquad
- (24 - 48 z + 64 z^2) \ln z \ln (1-z)
+( 16 - 32 z + 32 z^2) \ln^2 (1-z)
\nonumber \\ && \qquad
+(8 - 16 z +32 z^2)\ln^2 z
-( 28 - 96 z + 80 z^2) \ln (1-z) 
\nonumber \\ && \qquad
+( 8 - 48 z + 80z^2)
\ln z + 36 - 68z + 16 z^2\}
\nonumber \\ && \qquad
+ C_A T_f \{ - ( 16 + 32 z + 32 z^2)[ {\rm Li}_2(-z) 
+ \ln z \ln (1+z)] 
\nonumber \\ && \qquad
+(16 + 64 z) {\rm Li}_2(1-z)
-( 16 + 32 z^2) \zeta(2) 
\nonumber \\ && \qquad
+ ( 96 z - 32 z^2) \ln z \ln (1-z)
+(8 - 16 z + 16 z^2) \ln^2(1-z)
\nonumber \\ && \qquad
-(16 + 48 z) \ln^2 z +\Big( \frac{32}{3z} - 8 +160 z 
- \frac{536}{3} z^2\Big) \ln(1-z)
\nonumber \\ && \qquad
-(192 z - 200 z^2) \ln z +\frac{208}{9z} - \frac{220}{3}
- \frac{368}{3}z + \frac{1628}{9}z^2\}\Big] \ln\frac{Q^2}{m^2}
\nonumber \\ && \qquad
+ C_A T_f \Big[ \{ ( 16 - 32 z + 32 z^2) \ln(1-z)
+(16 + 64 z) \ln z +\frac{32}{3z} + 8 
\nonumber \\ && \qquad
+64 z - \frac{248}{3} z^2 \} \ln \frac{Q^2}{m^2}
\nonumber \\ && \qquad
+(16 + 64 z) {\rm Li}_2(1-z) -(16 - 32 z + 32 z^2) \zeta(2)
\nonumber \\ && \qquad
+ ( 96 z - 32 z^2) \ln z \ln (1-z) +( 16 - 32 z + 32 z^2)\ln^2(1-z)
\nonumber \\ && \qquad
-(8+32 z)\ln^2 z 
+\Big( \frac{32}{3z} - 8 + 192 z - \frac{632}{3} z^2 \Big) \ln (1-z)
\nonumber \\ && \qquad
-\Big(8 + 256 z - \frac{248}{3} z^2\Big) \ln z
+ \frac{16}{3z} - \frac{172}{3} - \frac{968}{3}z + \frac{1124}{3} z^2\Big]
\ln \frac{m^2}{\mu^2}
\nonumber \\ && \qquad
+ C_F T_f \Big[ 16 ( 1+z)^2 \Big( - 4 {\rm S}_{1,2}(-z)
- 4 \ln(1+z) {\rm Li}_2(-z) 
\nonumber \\ && \qquad
- 2 \zeta(2) \ln(1+z)
-2 \ln z \ln^2(1+z) + \ln^2 z \ln(1+z) - \frac{3}{2} {\rm Li}_2(1-z)\Big)
\nonumber \\ && \qquad
+ 8 (1-z)^2\Big( - 4{\rm S}_{1,2}(1-z)
- 3 {\rm Li}_3(1-z) + 12 {\rm Li}_3(-z) + \ln^3(1-z)
\nonumber \\ && \qquad
-\frac{5}{2} \ln z \ln^2(1-z)
+2\ln^2 z \ln (1-z) + \ln(1-z) {\rm Li}_2(1-z) 
\nonumber \\ && \qquad
- 4 \ln z {\rm Li}_2(-z) + 4 \zeta(2) \ln z - \frac{1}{3}\ln^3 z 
+3 \ln z \ln(1 - z)\Big)
\nonumber \\ && \qquad
+ 8 z^2 \Big( 4 {\rm S}_{1,2}(1-z) - {\rm Li}_3(1-z) +  \ln^3(1-z)
\nonumber \\ && \qquad
-\frac{7}{2} \ln z \ln^2(1-z)
+ 4 \ln^2 z \ln (1-z) - \ln(1-z) {\rm Li}_2(1-z) 
\nonumber \\ && \qquad
- 4 \zeta(2) \ln (1-z) 
+ 4 \ln z {\rm Li}_2( 1-z)
+ 4 \zeta(2) \ln z - \ln^3 z + 11 {\rm Li}_2(1-z)\Big)
\nonumber \\ && \qquad
+ 128 z {\rm Li}_3(-z) + (112 - 96 z
+ 192 z^2 ) \zeta(3) - ( 112 z - 144 z^2) \ln z \ln(1 - z)
\nonumber \\ && \qquad
 + \Big( \frac{16}{15z^2} + 96 + \frac{128}{3}z  + \frac{192}{5} z^3\Big)
[ {\rm Li}_2(-z) + \ln z \ln (1+z) ]
\nonumber \\ && \qquad
+( 48 - \frac{208}{3} z + 104 z^2 + \frac{192}{5} z^3 ) \zeta(2)
-( 22 - 88 z + 84 z^2) \ln^2(1-z)
\nonumber \\ && \qquad
-\Big(4 - \frac{8}{3} z + 52 z^2 +\frac{96}{5}z^3\Big) \ln^2 z 
+(28 -132 z + 96 z^2) \ln (1-z)
\nonumber \\ && \qquad
- \Big( \frac{16}{15 z} + \frac{712}{15} - \frac{136}{5} z 
+ \frac{672}{5} z^2 \Big) \ln z + \frac{16}{15z} - \frac{904}{15}
%
%
+ \frac{68}{5} z + \frac{328}{5} z^2 \Big]
\nonumber \\ && \qquad
+ C_A T_f \Big[ 16( 1 + 2z + 2 z^2) \Big( {\rm Li}_3( \frac{1-z}{1+z}) 
- {\rm Li}_3( - \frac{1-z}{1+z}) 
\nonumber \\ && \qquad
- \ln z \ln (1-z)
\ln (1+z) - \ln(1-z) {\rm Li}_2(-z) 
\nonumber \\ && \qquad
+ \frac{3}{4} \ln^2 z \ln (1+z)
+ \frac{3}{2} \ln z {\rm Li}_2(-z)\Big)
\nonumber \\ && \qquad
+ 8 ( 1 + 2z ) [ 5 {\rm S}_{1,2}(1-z) + 2 {\rm S}_{1,2}(-z)
- 3 {\rm Li}_3(-z) + \ln z \ln^2(1+z) 
\nonumber \\ && \qquad
+ 2\ln(1+z) {\rm Li}_2(-z) 
+ \zeta(2) \ln (1+z)]
\nonumber \\ && \qquad
+ 32 z [ 2 {\rm S}_{1,2}(1-z) + \ln z {\rm Li}_2(1-z)]
\nonumber \\ && \qquad
- ( 16 + 128 z) {\rm Li}_3(1-z) + 8 z^2[\ln^2 z \ln(1+z)
\nonumber \\ && \qquad
-2 \ln z {\rm Li}_2(-z)]
+(48 z - 16 z^2) \ln z \ln^2(1-z) 
\nonumber \\ && \qquad
-( 8 + 64 z - 16 z^2)\ln^2 z \ln(1-z)
+( 16 + 64 z ) \ln(1-z) {\rm Li}_2(1-z) 
\nonumber \\ && \qquad
- ( 40 - 48 z + 64 z^2) \zeta(2) \ln(1-z)
+\Big(\frac{16}{3} + 16 z \Big) \ln^3 z 
\nonumber \\ && \qquad
- ( 16 + 160 z - 32 z^2) \zeta(2) \ln z
-( 12 + 56 z + 8z^2) \zeta(3)
\nonumber \\ && \qquad
-\Big(\frac{32}{3z} + 48 - 16 z - \frac{208}{3} z^2\Big)   
[{\rm Li}_2(-z) + \ln z \ln(1+z)] + \Big( \frac{64}{3z} + 20
\nonumber \\ && \qquad
- 64 z + \frac{80}{3}z^2 \Big) {\rm Li}_2(1-z)
-\Big( \frac{32}{z} + 4 + 208 z - \frac{796}{3} z^2\Big) \zeta(2)
\nonumber \\ && \qquad
+( 16 - 288 z + 292 z^2) \ln z \ln (1-z) + \Big( \frac{16}{3z}
-6 + 64 z - \frac{214}{3} z^2\Big) \ln^2(1-z)
\nonumber \\ && \qquad
+( 184 z - 114 z^2) \ln^2 z +\Big(\frac{208}{9z} - \frac{112}{3}
-\frac{860}{3} z + \frac{2996}{9}z^2\Big)
\ln (1-z)
\nonumber \\ && \qquad
+\Big( \frac{292}{3} + 332 z - \frac{5780}{9} z^2 \Big) \ln z 
+ \frac{80}{9z} + \frac{466}{9} + \frac{260}{9}z - \frac{878}{9} z^2\Big] \,.
\end{eqnarray}
The coefficient function corresponding to the Bethe-Heitler process (2.13)
reads (see (4.20), (4.21))
\begin{eqnarray}
&& H^{(2)}_{L,q}(z,Q^2,m^2)=
C_F T_f \Big[ \Big\{ - 32 z \ln z +\frac{32}{3z} - 32 + \frac{64}{3}z^2\Big\}
\nonumber \\ && \qquad
\times \{ \ln \frac{Q^2}{m^2} + \ln \frac{m^2}{\mu^2} \}
- 32 z[ {\rm Li}_2(1-z) + \ln z \ln(1-z) -\ln^2 z]
\nonumber \\ && \qquad
+\Big(\frac{32}{3z} - 32 + \frac{64}{3}z^2\Big)
 \ln(1-z) +(32 - 32 z - 64 z^2) \ln z
\nonumber \\ && \qquad
- \frac{32}{9z} + \frac{32}{3} - \frac{128}{3}z + \frac{320}{9}z^2\Big]  \,,
\end{eqnarray}
\begin{eqnarray}
 && H^{(2)}_{2,q}(z,Q^2,m^2)=
C_F T_f\Big[ \Big\{ 8 (1+z) \ln z + \frac{16}{3z} + 4 
- 4 z - \frac{16}{3}z^2\Big\}
\ln^2 \frac{Q^2}{m^2}
\nonumber \\ && \qquad
+ \Big\{[ 16 (1+z) \ln z +\frac{32}{3z} + 8 - 8 z - \frac{32}{3}z^2]
\ln \frac{Q^2}{m^2}
\nonumber \\ && \qquad
+ 8 ( 1+z)[ 2 {\rm Li}_2(1-z) + 2 \ln z \ln (1-z) - \ln^2 z]
\nonumber \\ && \qquad
+\Big( \frac{32}{3z} + 8 - 8z - \frac{32}{3} z^2\Big)  \ln(1-z) 
\nonumber \\ && \qquad
- \Big( 8 + 40 z - \frac{32}{3}z^2\Big) \ln z +\frac{16}{3z} - 
\frac{160}{3} +\frac{16}{3} z + \frac{128}{3}z^2\Big\}
\ln\frac{m^2}{\mu^2}
\nonumber \\ && \qquad
+ \{ 16 ( 1+z) [ {\rm Li}_2(1-z) + \ln z \ln (1-z) - \ln^2 z]
\nonumber \\ && \qquad
+(\frac{32}{3z} + 8 - 8 z - \frac{32}{3}z^2) \ln(1-z)
+ 32 z^2 \ln z + \frac{208}{9z} - \frac{208}{3} 
\nonumber \\ && \qquad
+ \frac{160}{3}z - \frac{64}{9} z^2 \} \ln \frac{Q^2}{m^2}
\nonumber \\ && \qquad
+(1+z) \Big( 32 {\rm S}_{1,2}(1-z) - 16 {\rm Li}_3(1-z) + 8 \ln z \ln^2(1-z)
\nonumber \\ && \qquad
-16 \ln^2 z \ln(1-z) + 16 \ln(1-z) {\rm Li}_2(1-z) 
- 32 \zeta(2) \ln z +\frac{16}{3}\ln^3 z\Big)
\nonumber \\ && \qquad
-\Big(\frac{32}{3z} + 32 + 32 z + \frac{32}{3}z^2\Big) [ {\rm Li}_2(-z) 
+ \ln z \ln(1+z)] 
\nonumber \\ && \qquad
+\Big(\frac{64}{3z} + 16 - 16 z
+\frac{32}{3} z^2 \Big) {\rm Li}_2(1-z)
-\Big(\frac{32}{z} + 16 + 16 z - \frac{64}{3}z^2\Big) \zeta(2)
\nonumber \\ && \qquad
+32 z^2 \ln z \ln (1-z)+\Big( \frac{16}{3z}+4 - 4 z - \frac{16}{3}z^2\Big)
 \ln^2(1-z)
\nonumber \\ && \qquad
+(40z - 16 z^2) \ln^2 z +\Big(\frac{208}{9z} - \frac{208}{3}
+ \frac{160}{3} z - \frac{64}{9} z^2 \Big) \ln(1-z) 
\nonumber \\ && \qquad
+ \Big( \frac{280}{3}
- 88 z - \frac{704}{9} z^2 \Big) \ln z 
\nonumber \\ && \qquad
+\frac{80}{9z} + \frac{304}{9} -
\frac{1216}{9} z + \frac{832}{9} z^2 \Big]   \,.
\end{eqnarray}
Finally we present the coefficient functions corresponding to the Compton 
process (2.13). They are given by eqs. (4.28), (4.29) which become equal to
\begin{eqnarray}
 && L^{{\rm NS},(2)}_{L,q}(z,Q^2,m^2)=
C_F T_f \Big[ \frac{16}{3}z \Big( \ln \frac{Q^2}{m^2} + \ln(1-z) - 2\ln z\Big)
\nonumber \\ && \qquad\qquad
+ \frac{16}{3}  - \frac{400}{18}z\Big] \,,
\end{eqnarray}
\begin{eqnarray}
 && L^{{\rm NS},(2)}_{2,q}(z,Q^2,m^2)=
C_F T_f \Big[  \frac{4}{3} \left ( \frac{1+z^2}{1-z} \right )
 \ln^2 \frac{Q^2}{m^2}
+ \Big\{ \frac{1+z^2}{1-z} \Big( \frac{8}{3} \ln (1-z)
\nonumber \\ && \qquad
- \frac{16}{3} \ln z - \frac{58}{9} \Big)
+ \frac{2}{3} + \frac{26}{3} z \Big\} \ln \frac{Q^2}{m^2}
\nonumber \\ && \qquad
+ \Big( \frac{1+z^2}{1-z} \Big) \Big ( - \frac{8}{3} {\rm Li}_2(1-z)
- \frac{8}{3} \zeta(2) - \frac{16}{3} \ln z \ln (1-z) 
\nonumber \\ && \qquad
+ \frac{4}{3} \ln^2(1-z) 
+ 4 \ln^2 z - \frac{58}{9} \ln(1-z) + \frac{134}{9} \ln z
+ \frac{359}{27} \Big )
\nonumber \\ && \qquad
+\Big( \frac{2}{3} + \frac{26}{3}z\Big) \ln(1-z) 
- \Big( 2+\frac{46}{3}z \Big) \ln z + \frac{29}{9} - \frac{295}{9}z \Big] \,.
\end{eqnarray}
In the above expressions one should bear in mind that the singularity at
$z=1$ will never show up because of the kinematical constraint 
$z < Q^2/(Q^2+4m^2)$.
However after convoluting $L^{NS}_{2,q}$ by the parton densities the structure
function $F_2(z,Q^2,m^2)$ in (2.8) will diverge as $\ln^3(Q^2/m^2)$ in the limit
$Q^2 \gg m^2$. 
In the above limit the upper boundary in (2.8) $z_{max}\rightarrow1$
and the virtual gluon which decays into the heavy quark pair becomes soft. 
The soft gluon contribution which causes the cubic logarithm above has to be
added to the two-loop virtual gluon corrections calculated in appendix A of
\cite{rn} . The cubic logarithm is then cancelled. The final result will be
that in (D.8) the singular terms at $z=1$ have to be replaced by the
distributions $(\ln^k(1-z)/(1-z))_+$ defined by 
\begin{eqnarray}
\int_0^1\, dz \Big( \frac{\ln^k(1-z)}{1-z}\Big)_+ f(z) =
\int_0^1\, dz \Big( \frac{\ln^k(1-z)}{1-z}\Big) ( f(z) - f(1)) \,,
\end{eqnarray}
and one has to add the following
delta function contribution
\begin{eqnarray}
 && L^{{\rm NS},S+V,(2)}_{2,q}(z,Q^2,m^2)=C_FT_f\delta(1-z)\Big\{2\ln^2
\frac{Q^2}{m^2}
-\Big[\frac{32}{3}\zeta(2)+\frac{38}{3}\Big]\ln\frac{Q^2}{m^2}
\nonumber \\ && \qquad \qquad \qquad
+\frac{268}{9}\zeta(2) + \frac{265}{9}\Big\} \,.
\end{eqnarray}
\vfill
\newpage
%

%
\vfill
\newpage
\centerline{\bf \large{Figure Captions}}
\begin{description}
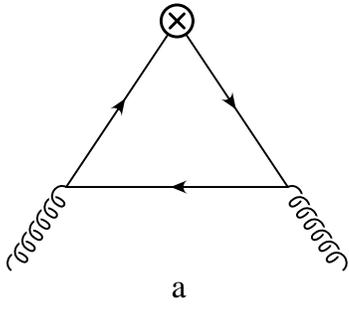
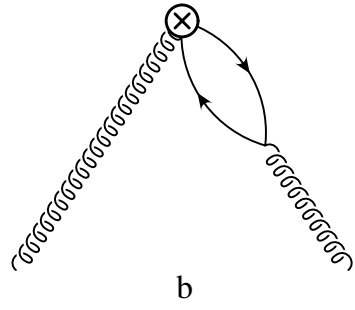
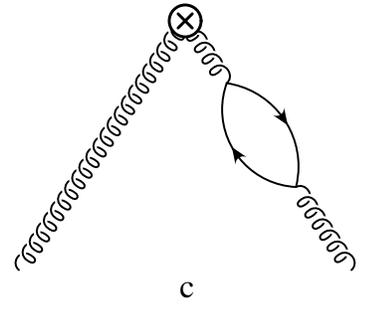
\item[Fig. 1.]
a,b : One-loop graphs contributing to the OME $A_{Qg}^{(1)}$.\\
c : Heavy quark 
($m_H^2 > Q^2$) loop contribution to $A_{gg}^{(1)}$. The solid line indicates 
the heavy quark $Q$.
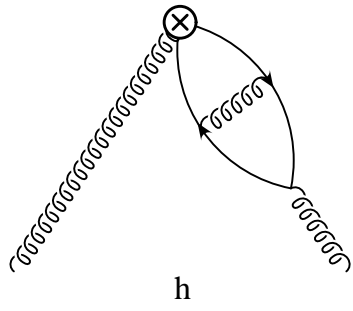
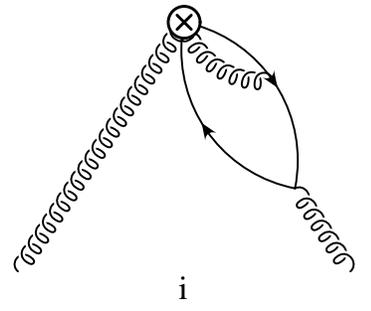
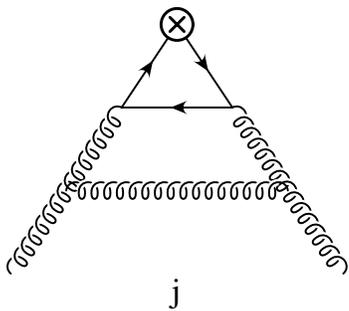
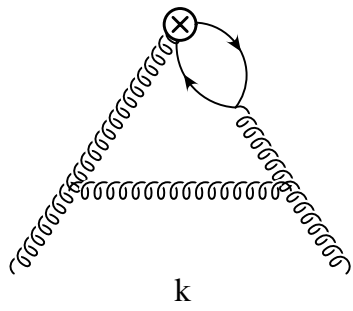
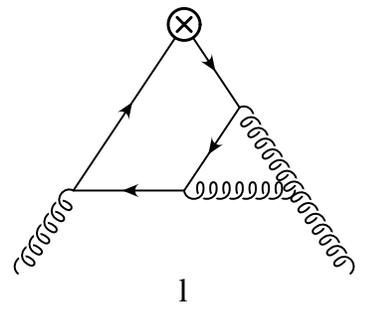
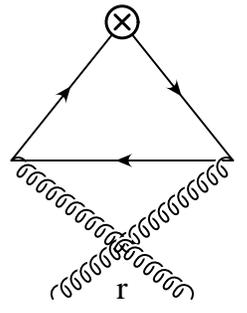
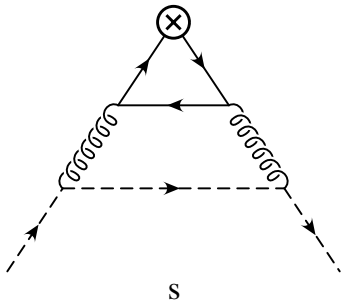
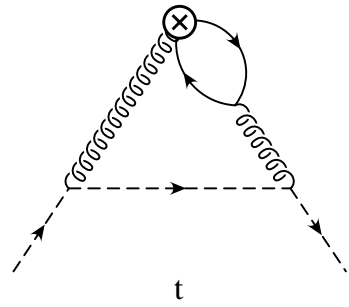
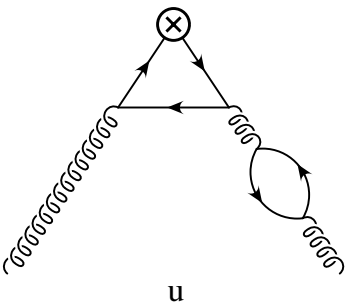
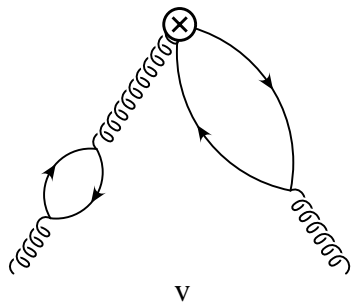
\item[Fig. 2.]
Two-loop graphs contributing to the OME $A_{Qg}^{(2)}$. The solid line
indicates the heavy quark $Q$. Diagrams $s$ and $t$ with the external ghosts 
(dashed line) are needed to cancel the unphysical polarizations which appear
in the sum of (3.43). Graphs $u$ and $v$ contain the external gluon self energy
with the heavy quark ($m_H^2 > Q^2 $) loop.
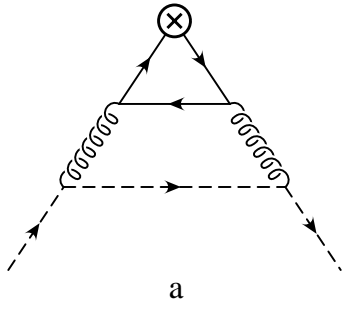
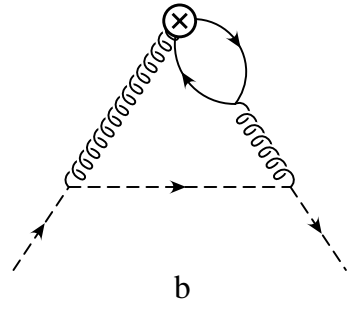
\item[Fig. 3.]
Two-loop graphs contributing to the OME $A_{Qq}^{{\rm PS},(2)}$. The solid line
represents the heavy quark $Q$ whereas the dashed line stands for the light
quark $q$.
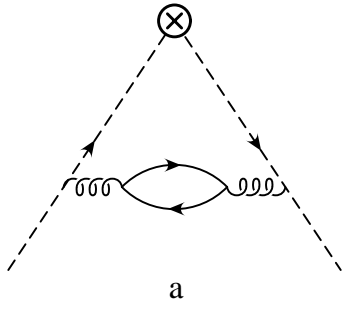
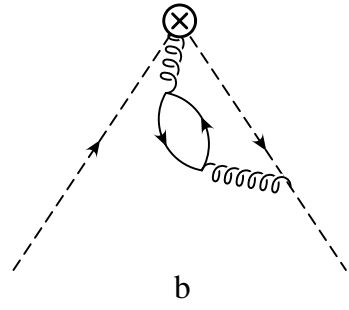
\item[Fig. 4.]
Two-loop graphs contributing to the OME $A_{qq}^{{\rm NS},(2)}$. The solid line
represents the heavy quark $Q$ whereas the dashed line stands for the light
quark $q$.
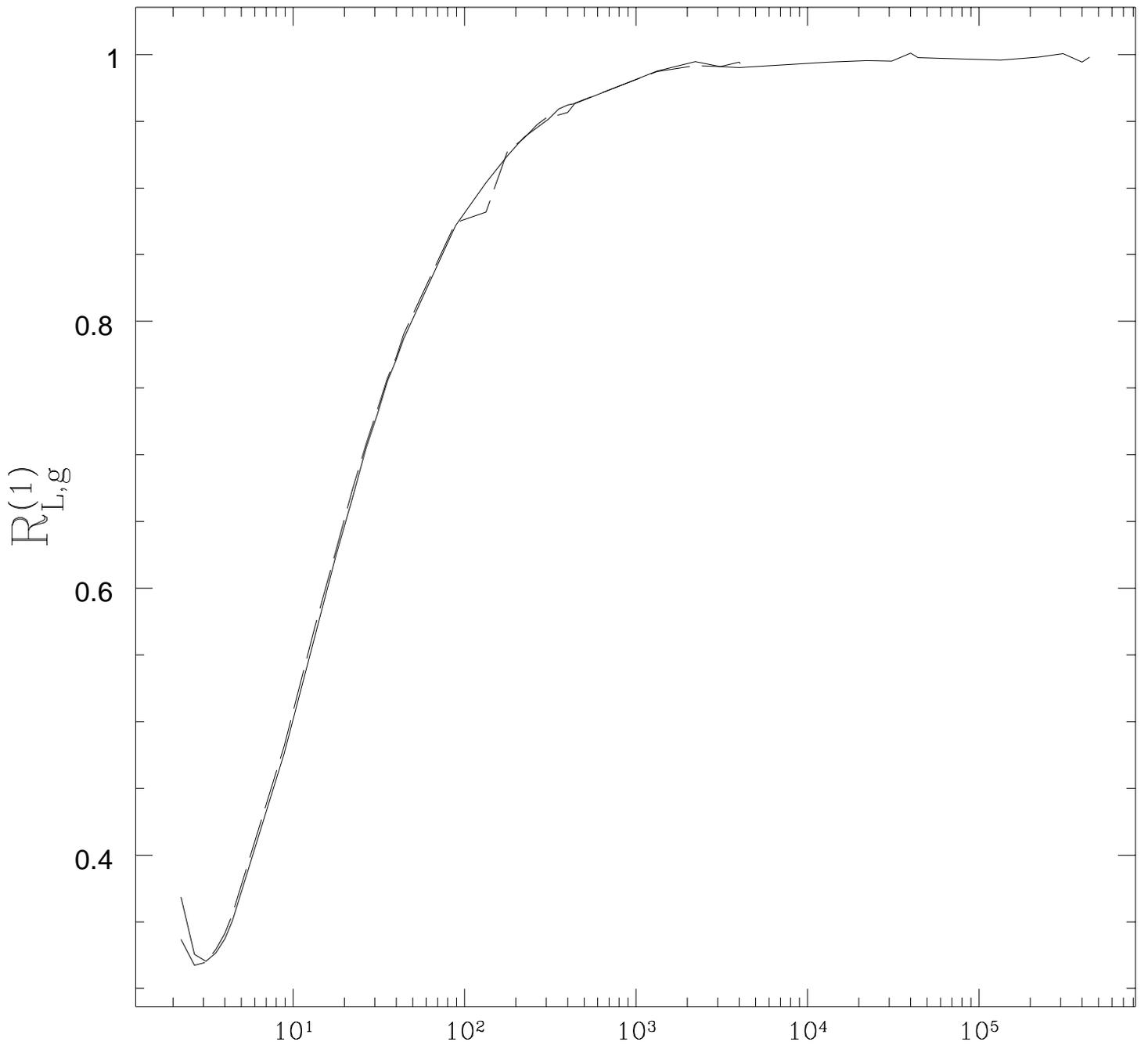
\item[Fig. 5.]
$R^{(1)}_{L,g}$ 
plotted as a function of $\xi$ for fixed
$z = 10^{-2}$  (solid line) and $z = 10^{-4}$ (dashed line).
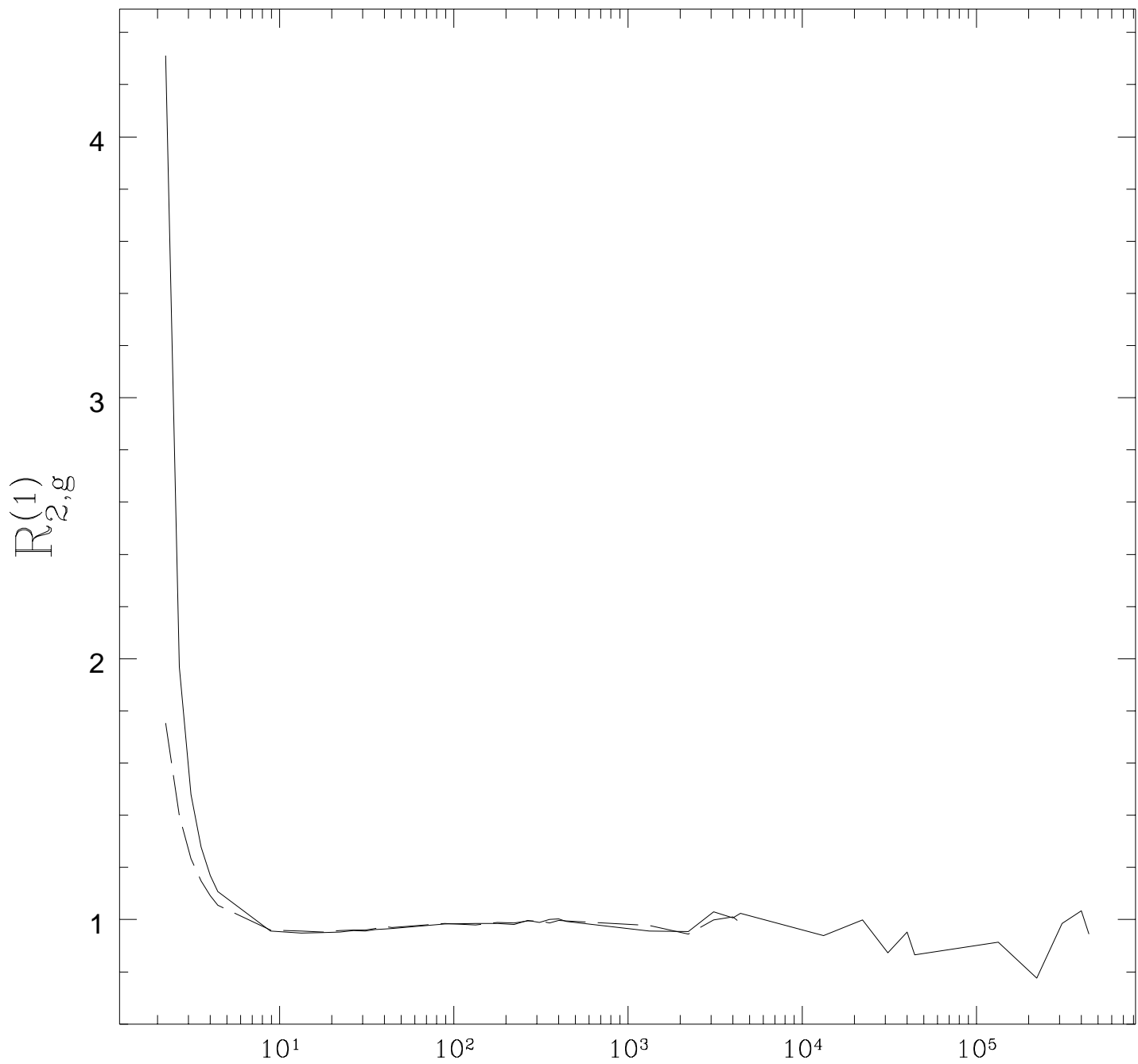
\item[Fig. 6.]
$R^{(1)}_{2,g}$
plotted as a function of $\xi$ for fixed
$z = 10^{-2}$  (solid line) and $z = 10^{-4}$ (dashed line).
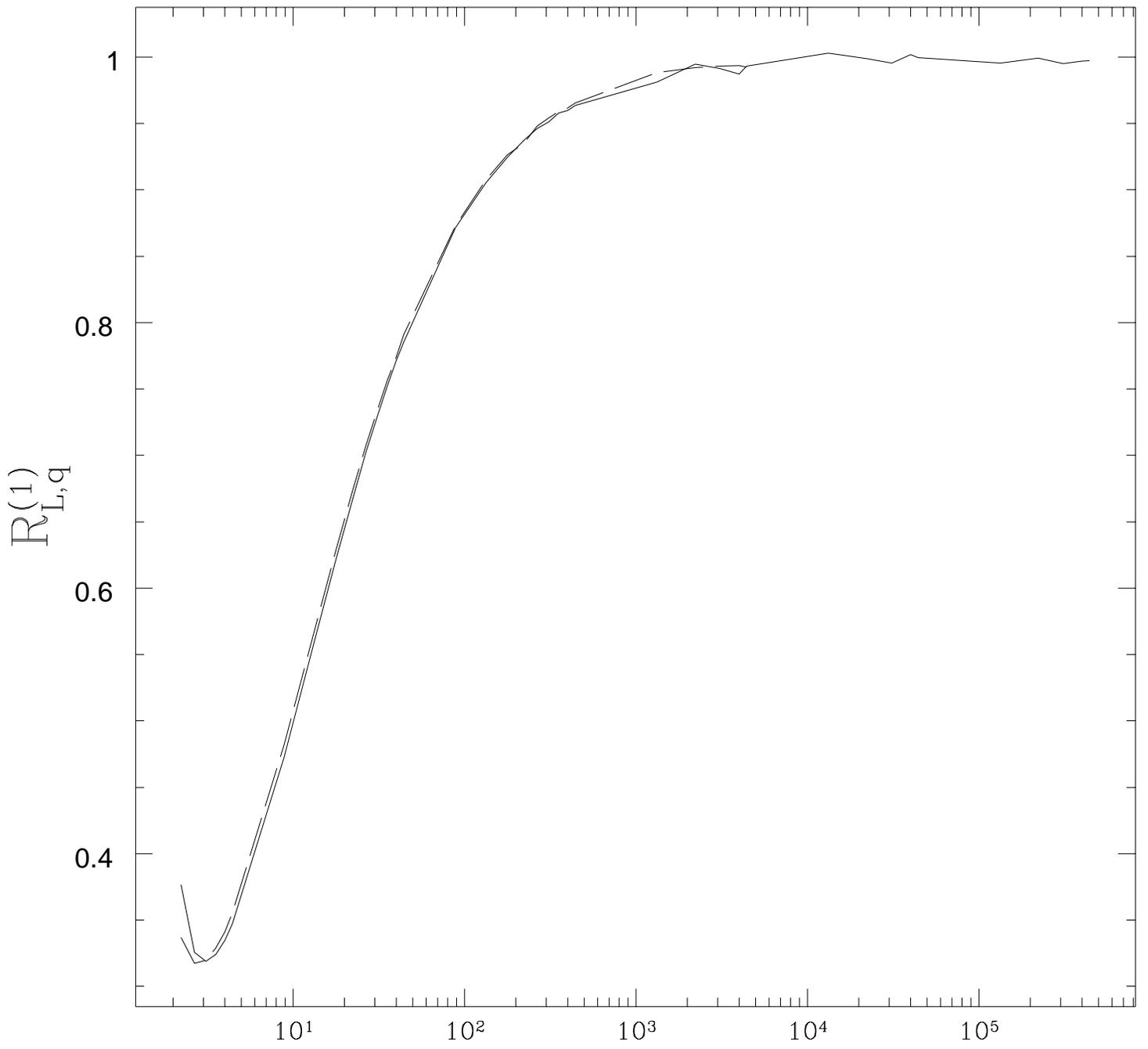
\item[Fig. 7.]
$R^{(1)}_{L,q}$
plotted as a function of $\xi$ for fixed
$z = 10^{-2}$  (solid line) and $z = 10^{-4}$ (dashed line).
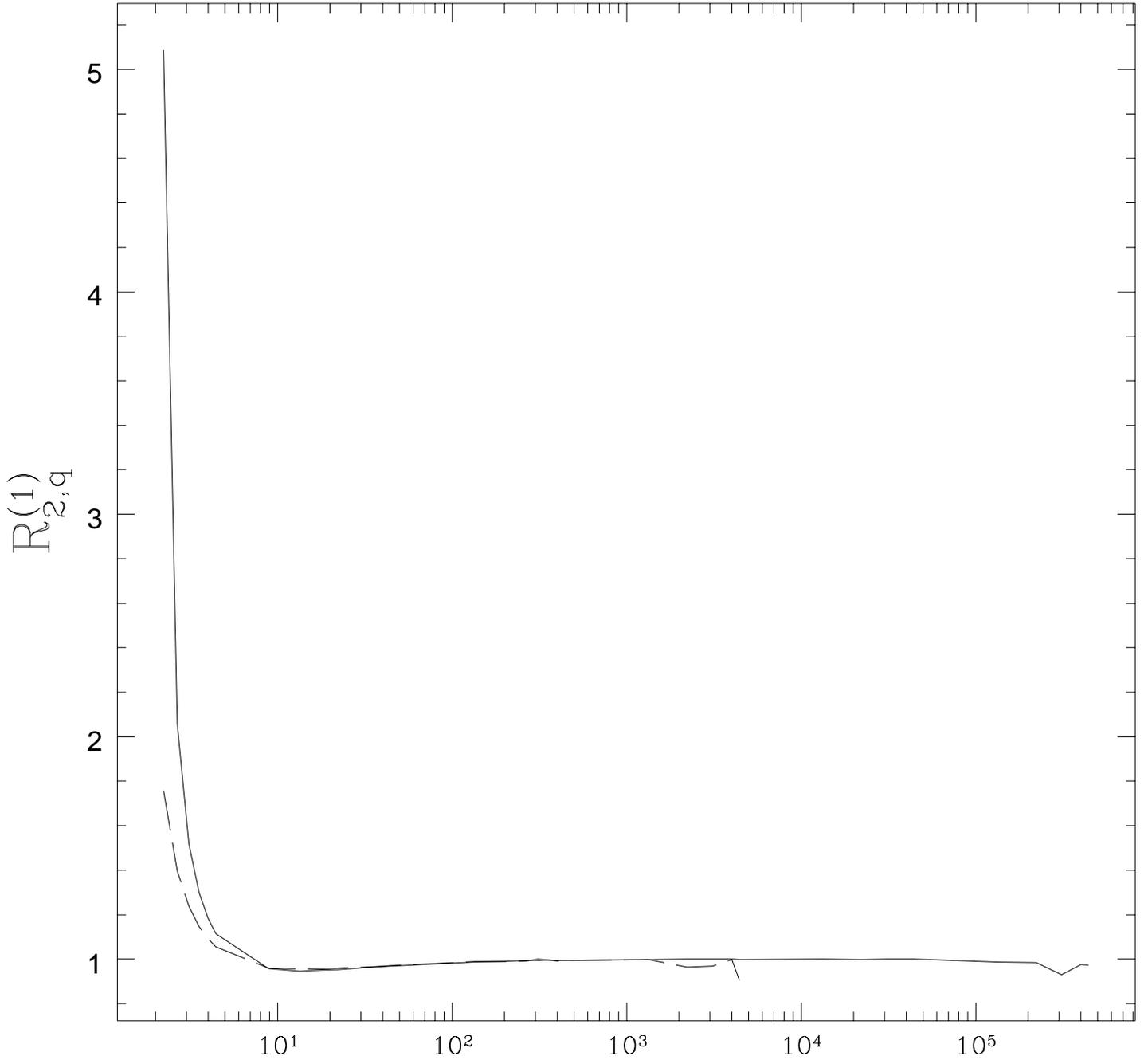
\item[Fig. 8.]
$R^{(1)}_{2,q}$
plotted as a function of $\xi$ for fixed
$z = 10^{-2}$  (solid line) and $z = 10^{-4}$ (dashed line).
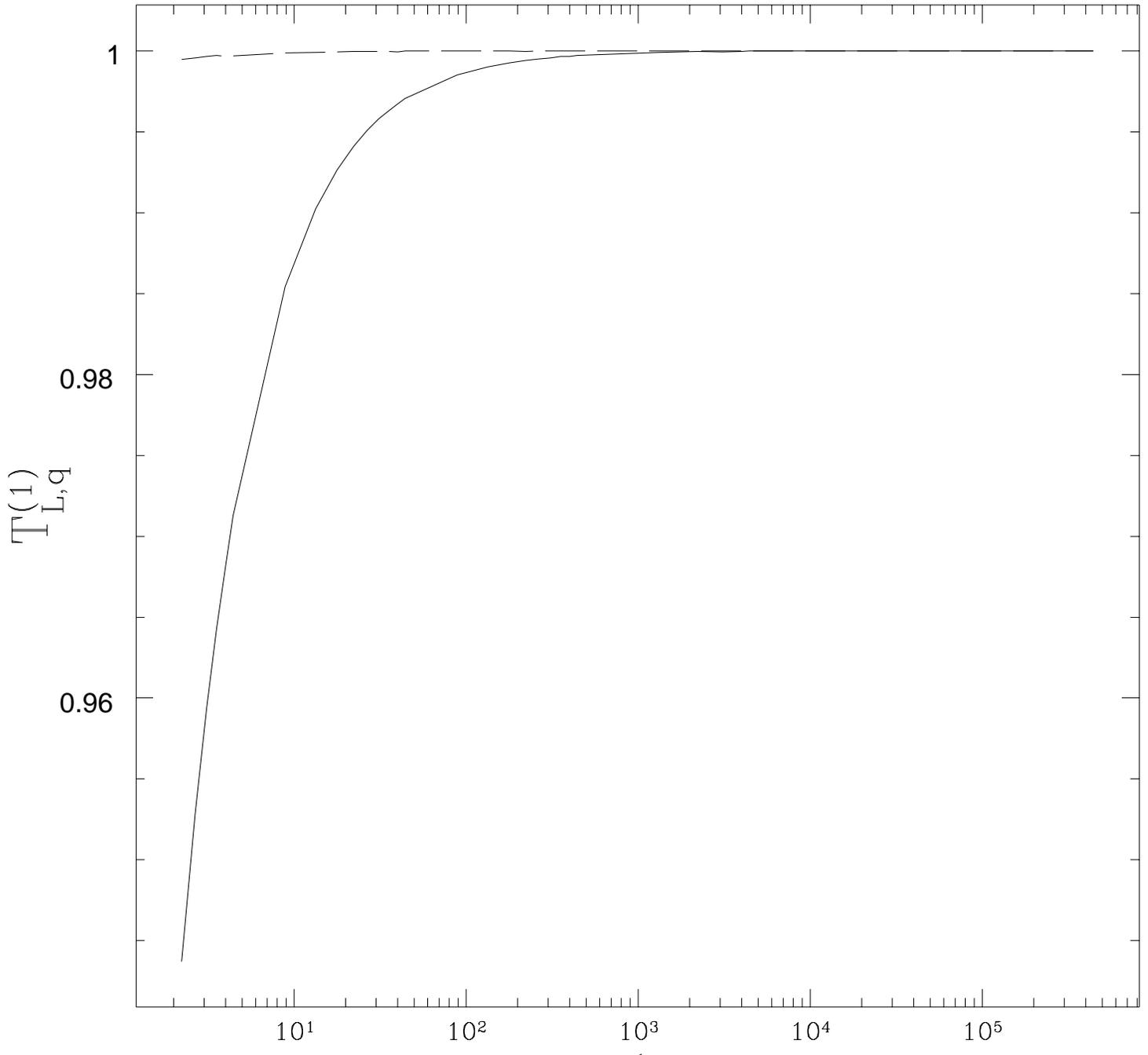
\item[Fig. 9.]
$T^{(1)}_{L,q}$
plotted as a function of $\xi$ for fixed
$z = 10^{-2}$  (solid line) and $z = 10^{-4}$ (dashed line).
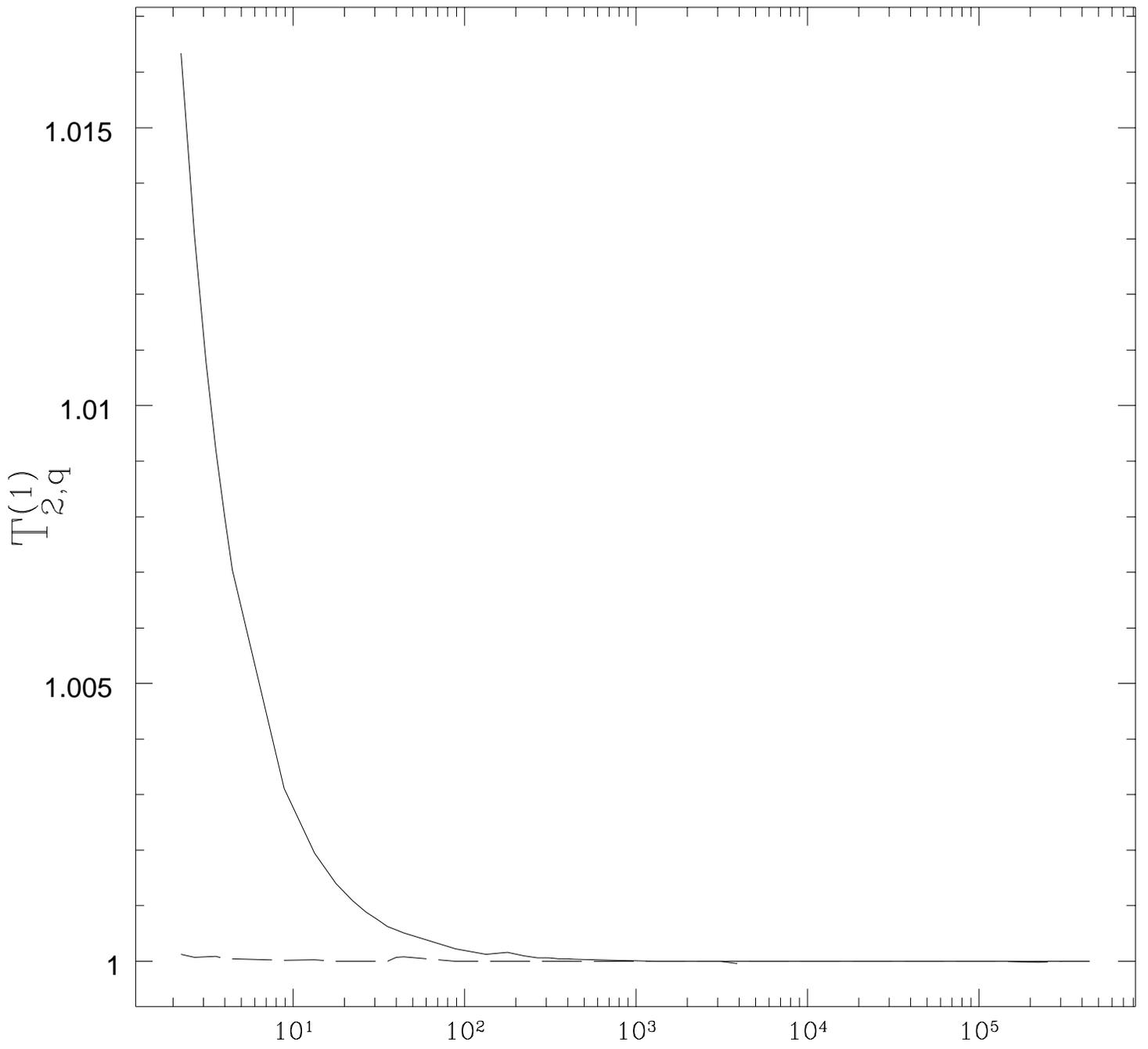
\item[Fig. 10.]
$T^{(1)}_{2,q}$
plotted as a function of $\xi$ for fixed
$z = 10^{-2}$  (solid line) and $z = 10^{-4}$ (dashed line).
\end{description}
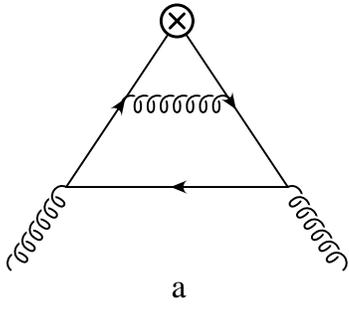
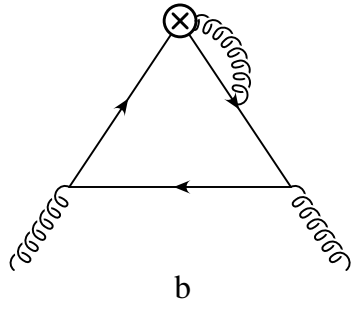
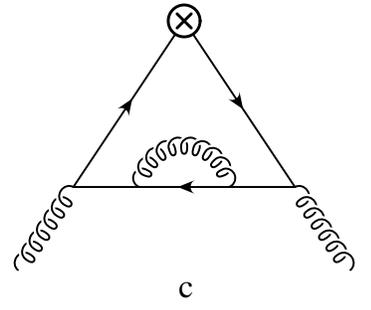
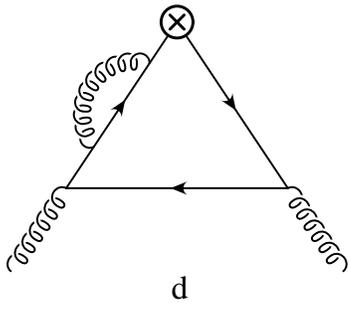
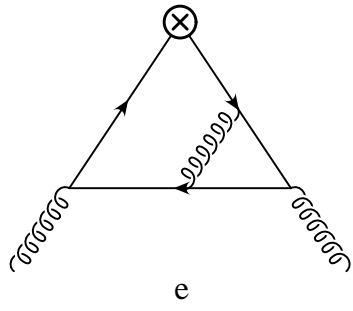
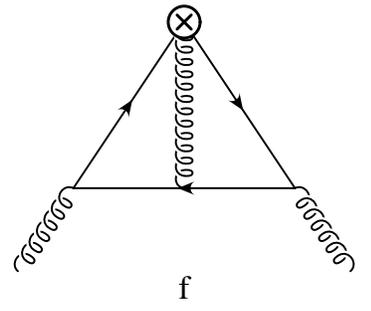
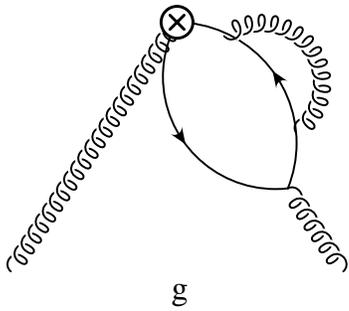
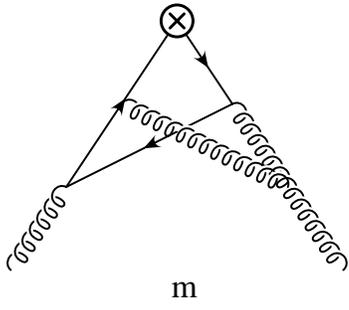
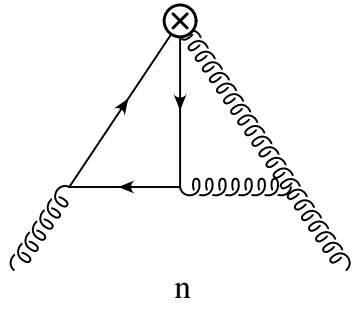
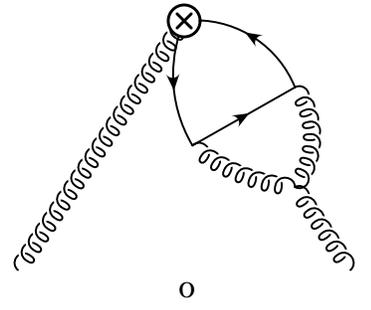
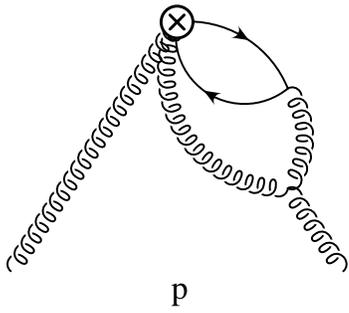
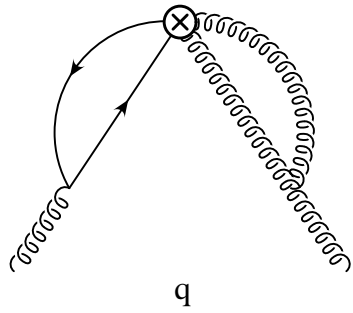
\end{document}